\documentclass[journal]{IEEEtran}
\input{decl}

\begin{document}

\title{Bandlimited signal reconstruction from\\
leaky integrate-and-fire encoding using POCS}

\author{Nguyen T. Thao,~\IEEEmembership{Member,~IEEE,}
Dominik Rzepka and Marek Mi\'skowicz,~\IEEEmembership{Senior Member,~IEEE} 
\thanks{N. T. Thao is with the Department of Electrical Engineering, The City College of New York, CUNY, New York, USA, email: tnguyen@ccny.cuny.edu.}
\thanks{D. Rzepka and M. Mi\'skowicz are with the Department of Measurement and Electronics, AGH University of Science and Technology, Krak\'ow, Poland, emails: drzepka@agh.edu.pl, miskow@agh.edu.pl}
\thanks{D. Rzepka and M. Mi\'skowicz were supported by the Polish National Center of Science under grants DEC-2017/27/B/ST7/03082 and DEC-2018/31/B/ST7/03874, respectively.}
}
\maketitle

\begin{abstract}
Leaky integrate-and-fire (LIF) encoding is a model of neuron transfer function in biology that has recently attracted the attention of the signal processing and neuromorphic computing communities as a technique of event-based sampling for data acquisition.
While LIF enables the implementation of analog-circuit signal samplers of lower complexity and higher accuracy simultaneously, the core difficulty of this technique is the retrieval of an input from its LIF-encoded output. In this article, we study this problem in the context of bandlimited inputs, by extracting the most abstract features of an LIF encoder as a generalized nonuniform sampler. In this view, the LIF output is seen as the transformation of the input by a known linear operator. We show that the signal reconstruction method of projection onto convex sets (POCS) converges to a weighted pseudo-inverse of this operator. This allows perfect recovery under uniqueness of reconstruction, minimum-norm reconstruction under incomplete sampling, as well as a noise shaping of time quantization that outperforms standard pseudo-inversion. On the practical side, a single iteration of the POCS method can be used to improve {\em any} estimate whose LIF samples are not consistent with those of the input, and a rigorous discrete-time implementation of this iteration is proposed that does not require a Nyquist-rate representation of the signals.
\end{abstract}

\begin{IEEEkeywords}
integrate and fire, leakage, bandlimited signals, nonuniform sampling, event-based sampling, time-encoding machine, time quantization, weighted pseudo-inverse, POCS, contraction.
\end{IEEEkeywords}

\section{Introduction}

\begin{figure}
\centerline{\hbox{\scalebox{0.6}{\includegraphics{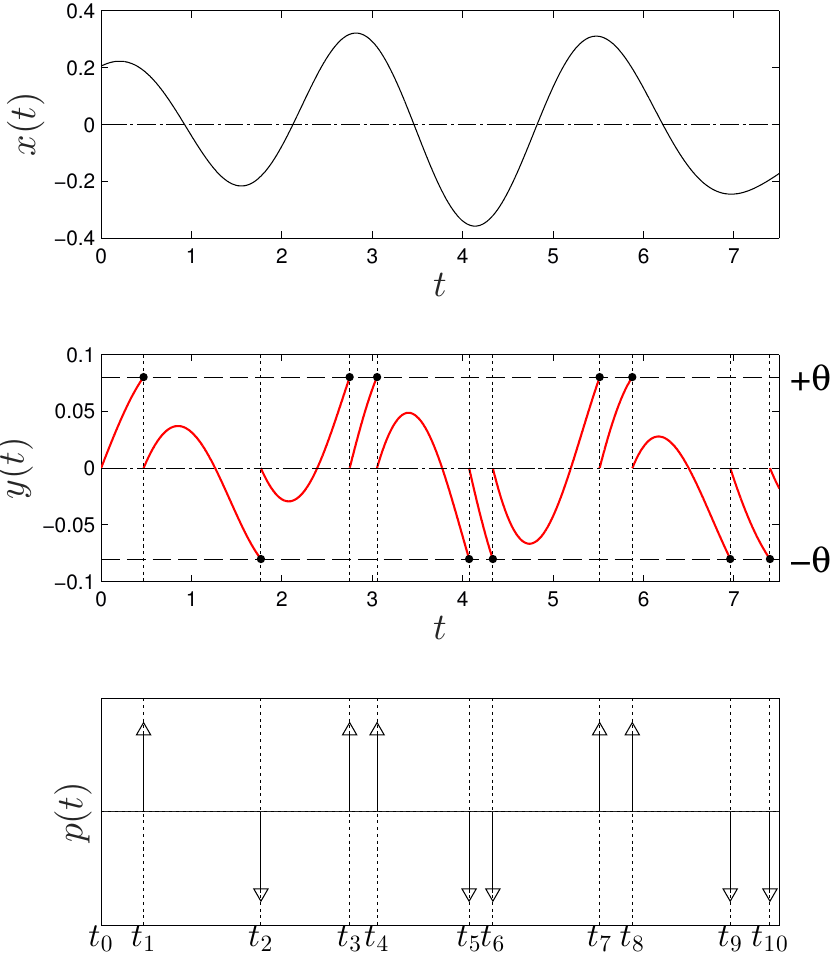}}}}
\caption{Encoding process of LIF (bipolar configuration with $c=0$).}\label{fig0}
\end{figure}

Integrate-and-fire (IF) encoding is a biologically-inspired model for mapping a continuous-time stimulus to a spike train. As opposed to traditional pulse-code modulation (PCM) in data conversion, which outputs a sequence of amplitude values, the input  information that an IF encoder provides is in the timing of its output spikes. It specifically fires an impulse whenever the integral of the stimulus reaches a given threshold (see Fig. \ref{fig0}). Time encoding is attracting more and more interest in data acquisition \cite{yoon2016lif,kianpour2019system,martinez2019delta,Adam19b,Alexandru20,Adam21,hilton2021time,Namman21} as the downscaling of semiconductor integration is increasing time precision while resulting in less amplitude accuracy \cite{9314920}. This is also part of the trend on event-based sampling \cite{Miskowicz2018} with more general objectives such as having acquisition activity dependent on input activity for power efficiency. What has made time encoding a difficult orientation is the non-trivial digital postprocessing it requires to convert the time information back to an explicit description of the input. This direction of research took off with the pioneering work of Lazar and T\'oth in \cite{Lazar04} which introduced an iterative algorithm that perfectly recovers bandlimited inputs encoded by asynchronous Sigma-Delta modulators (ASDM). This was simultaneously extended to IF encoding by Lazar in \cite{Lazar04b}. The present paper has two simultaneous goals:
\begin{enumerate}
\item Identify a general class of possible digital postprocessing methods for time encoding, by placing this problem in the broader framework of {\em generalized nonuniform sampling} \cite{eldar2005general}.
\item Develop and test the resulting postprocessing methods on leaky integrate-and-fire (LIF) encoding, which generalizes the application of ASDM and IF to time encoding.
\end{enumerate}

Leakage is a factor that was originally included in the IF model of a neuron to reflect the firing dynamics of the biological nervous systems with higher fidelity \cite{stein1965theoretical}. Interest in the processing of LIF encoded signals is expected to grow with the development of neuromorphic hardware \cite{Davies18,Bouvier19,Young19,furber2020spinnaker}. In data acquisition, leakage is simultaneously an artifact that is omnipresent in circuit integrators \cite{stata1967operational,norsworthy1996delta,nahvi2017schaum}.

At this stage, two iterative algorithms have been established for the bandlimited recovery of time encoded signals. The pioneering method of \cite{Lazar04} on ASDM-based time encoding requires some strict conditions of unique reconstruction. It was soon adapted to LIF encoding by Lazar in \cite{LAZAR2005401}. The alternative method of projection onto convex sets (POCS) was more recently introduced to ASDM encoding in \cite{Thao21a}, with the advantage of unconditional convergence including the situation of incomplete sampling, and the possibility of multiplierless FIR implementations. The basic contributions of the present paper are the following:
\begin{enumerate}[label=(\alph*)]
\item Extend the POCS method to LIF encoding, and compare its performance theoretically and numerically with Lazar's method.
\item Identify the fundamental algorithmic principle behind of these two methods to either understand their limitations or see their potentials for generalizations in nonuniform sampling.
\item With the invalidity of Fourier analysis due to time-varying processing, grow insight on the fundamental properties of these methods at the abstract level of linear algebra in Hilbert spaces.
\end{enumerate}

The article is organized as follows. We start the paper from the historical perspective of signal reconstruction in LIF encoding specifically. After giving the exact description of an LIF encoder in Section \ref{sec:setting} together with our signal assumptions and notation, we review some background in one-step bandlimited input estimation from LIF samples in Section \ref{sec:one-step}. This includes a deterministic technique proposed in \cite{Feichtinger12} and a classic statistical approach of neural research \cite{bialek91,rieke1999spikes,eliasmith2003neural}. In Section \ref{sec:improv}, we progressively introduce the POCS method as a basic technique to improve any bandlimited estimate that is not {\em consistent} with the input, i.e., that does not reproduce the same integral values as the input at the original firing instants. Numerical experiments show that the one-step methods of Section \ref{sec:one-step} are systematically improved in this process. Now, by alternating this local estimate improvement with bandlimitation, the POCS method  systematically converges to an estimate that is simultaneously bandlimited and consistent \cite{Combettes93,Bauschke96}. This leads to perfect reconstruction when the LIF output uniquely characterizes the input among the bandlimited signals. The numerical experiments also compare the convergence behavior of the POCS method and Lazar's iterative algorithm. While these two methods have a similar behavior in absence of leakage, Lazar's method appears to diverge at least when the leakage time constant is of the order of the Nyquist period even with an average density of firing instants that is 50\% above the Nyquist rate. The purpose of Sections \ref{sec:nonunif-samp}, \ref{sec:contract} and \ref{sec:noise} is to compare these two methods at a higher theoretical level. Section \ref{sec:nonunif-samp} first formalizes the LIF encoding process as providing generalized samples of the input in the form of its inner-products with given kernel functions. This abstract view was previously introduced in \cite{eldar2005general}. The collection of these samples is further presented as the transformation of the input through a linear operator $S$, which we call the {\em sampling operator}. As $S$ is in general not invertible, one naturally thinks of using a pseudo-inverse of it. This idea was previously raised in \cite{Wei05} in the form of pseudo-inversion of a single matrix. However, while this is a conceptually well-defined reconstruction scheme, no numerical method was suggested for this pseudo-inversion, in a context where the matrix size could be virtually infinite. We show in Section \ref{sec:nonunif-samp} that the POCS iteration limit actually achieves a weighted pseudo-inverse of $S$. This is done under weak assumptions on the encoder, allowing for example any integration kernel function, in place of the exponential leakage function of LIF encoding. We end this section with a reference to the work of \cite{Yen56} that considered for the first time sampling pseudo-inversion, in a context of incomplete point sampling. In Section \ref{sec:contract}, we show that both the POCS and Lazar's methods actually belong to the larger family of contraction algorithms \cite[\S1.2]{smart1980fixed} for solving a linear equation. This points their similarities as well as their differences in convergence at a more fundamental level, and gives more insight on the experimental results of Section \ref{sec:improv}. In Section \ref{sec:noise}, we study their behavior with respect to sampling noise in oversampling situations. This involves some non-standard analysis of time-varying noise shaping in nonuniform sampling in general. For practical evaluations, we focus in particular on errors due to time quantization, which is the counterpart of amplitude quantization in traditional data acquisition. We obtain the dramatic result that the weighted pseudo-inverse of the POCS method outperforms the standard pseudo-inverse in filtering this type of noise. As the POCS and Lazar's methods are theoretically defined on continuous-time signals, we finally present in Section \ref{sec:comp} a rigorous discretization of their iterations that does not involve Nyquist-rate resampling.

\section{Signal and system settings}\label{sec:setting}

\subsection{LIF system}

For a given input signal $x(t)$, the bipolar version of LIF outputs an impulse train sequence of the type
\begin{equation}\label{s}
p(t)=\textstyle\sum\limits_{n=1}^N\eps_n\delta(t-t_n)
\end{equation}
where $(t_n)_{1\leq n\leq N}$ is an increasing sequence of positive instants and $(\eps_n)_{1\leq n\leq N}$ is a sequence of signs. These two sequences are recursively defined by
\begin{equation}\label{ineq-inf}
\displaystyle t_n:=\min\Big\{t>t_{n-1}:\Big|\int_{t_{n-1}}^t e^{-\alpha(t-s)}\big(x(s)+c\big)\,\dif s\Big|
=\theta\Big\}
\end{equation}
and
\begin{equation}\label{equ-inf}
\int_{t_{n-1}}^{\raisebox{.3ex}{$\scriptstyle t_n$}} e^{-\alpha(t_n-s)}\big(x(s)+c\big)\,\dif s=\eps_n\theta
\end{equation}
starting from $t_0=0$, where $\alpha\geq0$, $c\geq0$ and $\theta>0$ are known constants. Fig. \ref{fig0} gives an illustration of this process with $c=0$. Unipolar LIF is the particular case where $x(t)+c\geq0$. Since all integral values are non-negative in this case, the absolute-value function can then be removed from \eqref{ineq-inf} and $\eps_n=+1$ for all $n$ in \eqref{equ-inf}.

For convenience, we will use the following notation throughout the paper,
\begin{eqnarray}
&\N:=\big\{1,{\cdots},N\},\nonumber\\
&\forall t\in\RR,\qquad\displaystyle A_\tau u(t):=\int_\tau^t e^{-\alpha(t-s)}u(s)\,\dif s,\quad \tau\in\RR,\label{A}\\
&x^c(t):=x(t)+c.\nonumber
\end{eqnarray}
Then, \eqref{ineq-inf} and \eqref{equ-inf} respectively take the simple forms of
\begin{equation}\label{ineq-inf2}
\displaystyle t_n=\min\big\{t>t_{n-1}:|A_{t_{n-1}}x^c(t)|=\theta\big\},\qquad n\in\N
\end{equation}
\begin{equation}\label{equ-inf2}
A_{t_{n-1}}x^c(t_n)=\eps_n\theta,\qquad n\in\N.
\end{equation}
Qualitatively speaking, LIF consists in detecting after each firing instant $t_n$ the next crossing of $A_{t_n}x^c(t)$ with the levels $\pm\theta$. LIF can thus be seen as a generalized version of level crossing sampling (LCS) \cite{Mark81,CTADC_2,Rzepka18}, where the thresholded signal is a dynamically changing integrated version of the input.

\subsection{Signal setting}\label{subsec:sig-setting}

All signals in this paper are assumed to be in the space $L^2(D)$ of real square-integrable functions on $D$, where $D$ is either $\RR$ or $[0,T]$. This space is equipped with the inner-product
\begin{equation}\label{inprod-def}
\langle u,v\rangle:=\int_D u(t) v(t)\,\dif t,
\end{equation}
which induces the norm $\|u\|:=\langle u,u\rangle^{1\!/2}$. Fourier transform is defined in both cases of domain $D$, with the difference that the frequency values are discrete and multiple of $2\pi/T$ when $D=[0,T]$ from the Fourier series of $T$-periodic signals.

For convenience, we assume that all bandlimited signals have Nyquist period 1 (which can always be achieved by a change of time unit). The baseband is then comprised of the frequencies $\omega\in[-\pi,\pi]$. We call $\calB$ the space of such signals. In the finite time case, we assume that $T$ is a multiple of the Nyquist period, and hence an integer. We denote by $\widetilde u(t)$ the bandlimited version of any signal $u(t)\in L^2(D)$. In the case $D=\RR$, we have
\begin{eqnarray}
&\widetilde u(t)=\varphi(t)*u(t)\nonumber\\[0.5ex]
\mbox{where}\qquad&\varphi(t):=\sin(\pi t)/(\pi t)&\qquad\qquad\label{phi}
\end{eqnarray}
and $*$ is the convolution operation. In the case $D=[0,T]$, we also have this expression assuming that $\varphi(t)$ is the periodic sinc function (or Dirichlet kernel) and $*$ is the circular convolution operation over $[0,T]$ (we assume that $T$ is an odd integer for the proper formation of the Dirichlet kernel).

The bounded domain $D=[0,T]$ is used in practice to designate the time window of signal acquisition. Given the decay of the sinc function, the $T$-periodic bandlimited signals asymptotically match the bandlimited signals of $L^2(\RR)$ as one looks at the time instants of $[0,T]$ that get away from the boundaries. But given that $T$ is in practice {\em virtually infinite} compared to the considered Nyquist period, the boundary effects are typically neglected. Meanwhile, $D=[0,T]$ theoretically makes $\calB$ of finite dimension (as a matter of fact, equal to $T$ given our assumptions), which allows us to define the situations of critical sampling and oversampling from nonuniform samples.

\section{One-step bandlimited estimation}\label{sec:one-step}

\subsection{Basic linear reconstruction}\label{subsec:basic}

In one-step bandlimited estimation, one looks at $x^c(t)$ as the input to the LIF encoder (in other words, one thinks of the constant component $c$ as part of the input). Using the output of the encoder, the goal is then to construct a bandlimited signal $u(t)$ that minimizes $\|u-x^c\|^2$, which we call the mean-squared error (MSE) of $u(t)$. A basic approach is to start searching for an estimate $u(t)$ that yields the same output as $x^c(t)$ through the same LIF encoder.  With the sequences $(t_n)\nN$ and $(\eps_n)\nN$ obtained from the encoding of $x(t)$, this amounts to requiring $u(t)$ to satisfy both \eqref{ineq-inf2} and \eqref{equ-inf2} recursively. Note that \eqref{ineq-inf2} amounts to inequality constraints while \eqref{equ-inf2} is an equality. The latter amounts to providing for each $n\in\N$ a sample of an affine transformation of $x(t)$. In the framework of generalized nonuniform sampling, we will simply be interested in estimates $u(t)$ that are consistent with the equality constraint of \eqref{equ-inf2}, i.e., such that
\begin{equation}\label{consistent}
\forall n\in\N,\qquad A_{t_{n-1}} u(t_n)=\eps_n \theta.
\end{equation}
The simplest way to obtain such an estimate is to take the signal
\begin{equation}\label{u0}
u_0(t):=\smallsum{n\in\N}\eps_n \theta\,\delta(t-t_n).
\end{equation}
Indeed, if one thinks of $\int_a^b v(s)\dif s$ as the integral of $v(s)$ from $a^+$ to $b^+$, then $u_0(t)$ is easily seen to satisfy \eqref{consistent}. But as $u_0(t)$ is not bandlimited, one adopts as final approximation of $x(t)$ its bandlimited version
$$\widetilde u_0(t)=\varphi(t)*u_0(t)=\smallsum{n\in\N}\eps_n \theta\,\varphi(t-t_n).$$
In this process, $\widetilde u_0(t)$ is likely to lose consistency. But we will keep this simple reconstruction as reference.

\subsection{Relation to the prior work of \cite{Feichtinger12}}

With $D=\RR$, the work of \cite{Feichtinger12} can be interpreted as starting from $u_0(t)$ as an initial estimate but proposing a different bandlimited transformation of it as final approximation of $x(t)$. The authors consider the global operator
$$\forall t\in\RR,\qquad Au(t):=A_{-\infty}u(t)=\int_{-\infty}^t e^{-\alpha(t-s)}u(s)\,\dif s$$
to take advantage of the following features:
\begin{itemize}
\item $A$ commutes with bandlimitation since it is a convolution operator,
\item $A$ is invertible of inverse $A^{-1}v(t)=v'(t)+\alpha\,v(t)$,
\item an estimate $u(t)$ that is 0 for all $t\leq0$ can be shown to satisfy \eqref{consistent} if and only if $Au(t_n)=Au_0(t_n)$ for all $n\in\N$.
\end{itemize}
Then, they propose to estimate $x(t)$ with a signal $\hat u_0(t)$ such that $A\hat u_0(t)$ is a bandlimited approximation of $Au_0(t)$. A downside of their approach, however, is that they specifically take the Nyquist-rate aliased version of $Au_0(t)$
\begin{equation}\label{Ahu0}
A\hat u_0(t)=\smallsum{k\in\ZZ} Au_0(k)\,\varphi(t-k)
\end{equation}
under our setting of Nyquist period 1. While this enables the simple derivation of $\hat u_0(t)$ as $\sum_{k\in\ZZ} Au_0(k)\,\psi(t-k)$
where
$\psi(t):=A^{-1}\varphi(t)=\varphi'(t)+\alpha\,\varphi(t)$, the aliasing in \eqref{Ahu0} is expected to be significant as $Au_0(t)$ is typically non-bandlimited given its jumps at the instants $(t_n)\nN$. In contrast, $A\widetilde u_0(t)$ yields the aliasing-free relation
\begin{align*}
A\widetilde u_0(t)&=\varphi(t)*Au_0(t)
\end{align*}
since $A$ commutes with bandlimitation. Not only is $\widetilde u_0(t)$ substantially simpler to implement than $\hat u_0(t)$, but the numerical experiments of Section \ref{subsec:numexp} also confirm its higher accuracy as an estimate of $x^c(t)$.

\subsection{Optimal data-driven convolutional reconstruction}

Note that the bandlimited estimate $\widetilde u_0(t)$ is of the form
\begin{equation}\nonumber
u(t)=f(t)*p(t)
\end{equation}
where $p(t)$ is defined in \eqref{s} and $f(t)$ is specifically chosen to be $\theta\,\varphi(t)$. A basic direction is to look for better reconstructions in this larger family of signals. This has been a basic approach of neural research to estimate the stimulus of a neuron from its output spike train. As the exact transfer function of a neuron is unknown, this direction of research has been considered based on statistics of effective input-output pairs of the neuron \cite{bialek91} \cite[\S2.3]{rieke1999spikes}. This method has been explicitly applied on LIF in \cite[\S4.3.3]{eliasmith2003neural} as demonstration. Assuming that $x(t)$ is a random process and $p(t)$ is the resulting response of the LIF encoder, the goal is to find the function $f\subsmall{\mathrm{opt}}(t)$ that minimizes
\begin{equation}\label{expectation}
E\Big(\big\|f(t)*p(t)-x^c(t)\big\|^2\Big).
\end{equation}
This is a Wiener filtering problem. Calling $U(\omega)$ the Fourier transform of $u(t)$, it is known (see for example \cite[(7.3.2)]{woyczynski2010first}) that
\begin{equation}\label{Fw}
F\subsmall{\mathrm{opt}}(\omega)=\frac{E\big(S(\omega)^*X^c(\omega)\big)}{E\big(|S(\omega)|^2\big)},
\end{equation}
for all $\omega$ in the baseband, and $F\subsmall{\mathrm{opt}}(\omega)=0$ otherwise since $f\subsmall{\mathrm{opt}}(t)$ must obviously be bandlimited. This leads to the definition of the new estimate
\begin{equation}\label{ulin}
u\subsmall{\mathrm{opt}}(t):=f\subsmall{\mathrm{opt}}(t)*p(t)=\smallsum{n\in\N}\eps_n\,f\subsmall{\mathrm{opt}}(t-t_n).
\end{equation}
The MSE of $u\subsmall{\mathrm{opt}}(t)$ is of course expected to be smaller than that of $\widetilde u_0(t)$. This is at the price of having the input-output statistics of the system. But it will be interesting to know for reference how much MSE reduction can be achieved with this method, as done in the next section.

When $c>0$, note that $x^c(t)$ cannot be in $L^2(\RR)$. In this case, \eqref{expectation} and \eqref{Fw} are at least well defined with $D=[0,T]$. With the theoretical setting of $D=\RR$, \eqref{expectation} and \eqref{Fw} can still be used provided that the constant components are removed from the analysis, thus making $u\subsmall{\mathrm{opt}}(t)$ optimal only up to a constant component.

\subsection{Numerical experiments}\label{subsec:numexp}

\begin{figure}
\centerline{\hbox{\scalebox{0.63}{\includegraphics{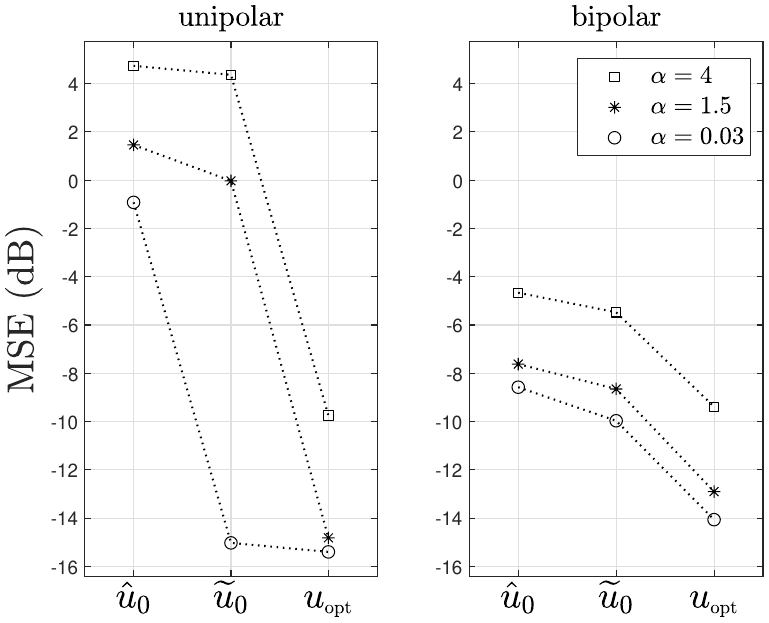}}}}
\caption{MSE of $\hat u_0(t)$,  $\widetilde u_0(t)$ and $u\subsmall{\mathrm{opt}}(t)$ for various leakage values $\alpha$ in the unipolar and bipolar cases.}\label{fig1}
\end{figure}

We show in Fig. \ref{fig1} the MSE results we got numerically for the bandlimited reconstruction estimates $\hat u_0(t)$, $\widetilde u_0(t)$ and $\widetilde u\subsmall{\mathrm{opt}}(t)$ with various leakage values $\alpha$ in the unipolar and bipolar cases. The numerical difficulty is to eliminate the boundary effects so that the true MSE from steady state can be extracted. We do this by working with bandlimited inputs that are periodic over 61 Nyquist periods. This enables the exact implementation of the sinc filter without boundary deviations. We adjust the threshold $\theta$ so that the number of firing instants per Nyquist period is in average equal to 1.5. We call this number the oversampling ratio. The MSE of each estimate is averaged over 100 zero-mean bandlimited inputs whose Nyquist-rate samples are drawn randomly and uniformly in $[-0.7,0.7]$. We use $c=1$ and $c=0$  in the unipolar and the bipolar cases, respectively. We take as 0 dB reference the averaged value of $\|x\|^2$. We derive the optimal filter $f\subsmall{\mathrm{opt}}(t)$ of linear reconstruction based on the statistics of 1000 bandlimited input thus randomly drawn. Given our periodic input setting, it is not possible to test the estimate $\hat u_0(t)$ in the absolute leakage-free case $\alpha=0$ since the operator $A$ does not converge. To simulate zero leakage, we set $\alpha$ to 0.03, which is close enough to 0 while allowing numerically a reliable convergence of $A$. Our observations are as follows:
\begin{enumerate}
\item Accurate signal reconstruction is more difficult as leakage increases.
\item $\widetilde u_0(t)$ is a better estimate than $\hat u_0(t)$ in all cases. The improvement is particularly strong in the leakage-free unipolar case.
\item As expected, the estimate $u\subsmall{\mathrm{opt}}(t)$ is in all cases as good as or better than $\widetilde u_0(t)$.
\item The experiment tends to show that $\widetilde u_0(t)$ is optimal as a convolutional reconstruction with the leakage-free unipolar configuration. Indeed, not only $\widetilde u_0(t)$ is close to $u\subsmall{\mathrm{opt}}(t)$ in MSE in the unipolar case with $\alpha=0.03$, but $F\subsmall{\mathrm{opt}}(\omega)$ also appears to be numerically close to the constant $\theta$ in $[-\pi,\pi]$. We attribute its remaining difference with the ideal constant to boundary effects together with a leakage that is not exactly 0.
\item The estimates $\hat u_0(t)$ and $\widetilde u_0(t)$ are basically useless in the unipolar case with leakage $\alpha\geq1.5$ as their MSE's are of the same order of (if not larger than) the input's squared sum.
\end{enumerate}

\section{Reconstruction by POCS}\label{sec:improv}

The signal estimation methods presented until now remain of limited accuracy as can be seen in the results of Fig. \ref{fig0}. This can be explained by a limited use of the analytical information contained in \eqref{equ-inf2}. An early observation is that little is done in these methods to make the bandlimited estimates consistent with this information. This was already mentioned concerning $\widetilde u_0(t)$. Meanwhile, the estimate $u\subsmall{\mathrm{opt}}(t)$ only focused on a convolutional improvement of $\widetilde u_0(t)$ while the LIF encoder is neither linear nor time invariant. We show in this section that there is a systematic way to improve any estimate that is not consistent. The POCS method is an algorithm that combines this improvement mechanism with the bandlimitation requirement, to converge to an estimate that is both bandlimited and consistent with  iterative MSE reductions.

\subsection{LIF encoder as generalized sampler}

The starting point is to rigorously reformulate the signal problem in the Hilbert space $L^2(D)$. This leads us to look at $x(t)$ as the input to the LIF encoder while thinking of $c$ as a parameter of the encoder. In this context, the MSE of an estimate $u(t)$ will refer to $\|u-x\|^2$. Next,  \eqref{equ-inf2} is restated as
\begin{equation}\label{Ax}
\forall n\in\N,\quad A_{t_{n-1}} x(t_n)=\theta_n:=\eps_n\theta-A_{t_{n-1}} c(t_n)
\end{equation}
where $c(t):=c$ for all $t$. Let us introduce the notation
$$I_n:=[t_{n-1},t_n)\quad\mbox{and}\quad\Delta t_n:=t_n-t_{n-1},\qquad n\in\N.$$
From \eqref{A}, \eqref{Ax} can be rewritten as
\begin{equation}\label{equ-inf3}
\forall n\in\N,\qquad \langle h_n,x\rangle=\theta_n
\end{equation}
where $\langle\cdot,\cdot\rangle$ was defined in \eqref{inprod-def},
\begin{equation}\label{hn}
h_n(t):=e^{-\alpha(t_n-t)}\,1_{I_n}(t)
\end{equation}
and $1_{I_n}(t)$ is the indicator function of the time interval $I_n$. In \eqref{equ-inf3}, $A_{t_{n-1}} x(t_n)$ is better seen as a linear functional mapping $x(t)$ into a single scalar value $\theta_n$ that fulfills the role of {\em sample} of $x(t)$, according to the generalized framework of sampling \cite{eldar2005general}. The function $h_n(t)$ is the associated {\em sampling kernel function}. The sample value $\theta_n$ is explicitly given by
$$\theta_n=\eps_n\theta-c\int_{t_{n-1}}^{t_n} e^{-\alpha(t_n-s)}\dif s
=\eps_n\theta-\midfrac{c}{\alpha}(1-e^{-\alpha\Delta t_n}).$$
Now, another contribution of this presentation is the orthogonality that results from  $\langle\cdot,\cdot\rangle$. Induced by this inner-product, the norm $\|\cdot\|$ satisfies the Pythagorean theorem which, in a stronger form, lies in the equivalence
\begin{equation}\label{Pyth}
\langle u,v\rangle=0\qquad\Leftrightarrow\qquad\|u+v\|^2=\|u\|^2+\|v\|^2
\end{equation}
for any $u(t),v(t)\in L^2(D)$.
This will be the key to MSE reductions as seen in the next section.

\subsection{Set theoretic estimation}

Another way to state the property of \eqref{equ-inf3} is to say that $x(t)$ belongs to the set
\begin{equation}\label{C}
\calC=\calC(\btheta):=\Big\{u(t)\in L^2(D):\forall n\in\N,~\langle h_n,u\rangle=\theta_n\Big\}
\end{equation}
where $\btheta$ symbolizes the sequence $(\theta_n)_{n\in\N}$. We call the elements of $\calC$ the estimates of $L^2(D)$ that are consistent with the sampling of $x(t)$ in the sense of \eqref{equ-inf3}. We will use the short notation $\calC$ when there is no ambiguity about the considered sequence $\btheta$. We alluded in Section \ref{subsec:basic} to the difficulty to find a bandlimited signal that is simultaneously consistent. Whenever an estimate $u(t)$ is in $L^2(D)$ but not in $\calC$, there is in fact a systematic procedure to reduce its MSE. This is due to the outstanding property that $\calC$
is an affine subspace of $L^2(D)$ (i.e., a translated linear subspace) which is moreover closed (since it is of finite codimension). As a generalization from closed linear subspaces, every signal $u(t)\in L^2(D)$ has an orthogonal projection $P_\calC u(t)$ onto $\calC$ in the sense of $\langle\cdot,\cdot\rangle$, which is the unique element of $\calC$ such that
\begin{equation}\label{orthoproj}
\forall v(t)\in\calC,\qquad\big\langle u-P_\calC u,\,P_\calC u-v\big\rangle=0.
\end{equation}
Due to \eqref{Pyth}, this is equivalent to
\begin{equation}\label{Pyth2}
\forall v(t)\in\calC,\qquad\|u-v\|^2=\|u-P_\calC u\|^2+\|P_\calC u-v\|^2.
\end{equation}
This gives the alternative characterization that $P_\calC u(t)$ is the element $v(t)$ of $\calC$ that minimizes $\|u-v\|$, i.e., that is closest to $u(t)$ in the MSE sense. As $x(t)\in\calC$, \eqref{Pyth2} also implies with $v(t)=x(t)$ that
$$\|P_\calC u-x\|\leq\|u-x\|$$
with a strict inequality when $u(t)\notin\calC$ since $\|u-P_\calC u\|>0$ in \eqref{Pyth2}.
In other words, $P_\calC u(t)$ is a better estimate of $x(t)$ than $u(t)$.

\subsection{Projection implementation}

It remains to find the explicit expression of $P_\calC$. What makes its derivation easy is the outstanding property that $(h_n(t))\nN$ is an {\em orthogonal} family of functions in $L^2(D)$ since their time supports do not overlap.
\line
\begin{proposition}\label{prop:PC}
Let $\calC=\calC(\btheta)$ be defined by \eqref{C} for any {\em orthogonal} family of functions $(h_n(t))\nN$ of $L^2(\RR)$. Then for all $u(t)\in\L^2(\RR)$,
\begin{equation}\label{PCu-explicit}
P_\calC u(t)=u(t)+\smallsum{n\in\N}(\theta_n-\langle h_n,u\rangle)\,\midfrac{h_n(t)}{\|h_n\|^2}.
\end{equation}
\end{proposition}
\line
\begin{IEEEproof}
Let $q(t)$ be the right hand side of \eqref{PCu-explicit}. Since $\langle h_m,h_n\rangle=0$ for any distinct $m,n\in\N$, then $\langle h_m,q\rangle=\langle h_m,u\rangle+(\theta_m-\langle h_m,u\rangle)=\theta_m$ for any $m\in\N$. So $q(t)\in\calC$. For any $v(t)\in\calC$, $\langle h_n,q-v\rangle=\theta_n-\theta_n=0$ for all $n\in\N$. Since $u(t)-q(t)=\sum\nN\alpha_n h_n(t)$ for some coefficients $\alpha_n$, then $\langle u-q,q-v\rangle=0$. Thus $q(t)=P_\calC u(t)$.
\end{IEEEproof}

\subsection{POCS algorithm}\label{subsec:POCS}

We can see from \eqref{PCu-explicit}, that the correcting term of $P_\calC u(t)$ is a piecewise exponential function, which is non-bandlimited. By considering the filtered version \begin{equation}\label{bandlimit-proj}
M\subP u(t):=\varphi(t)*P_\calC u(t)
\end{equation}
where $\varphi(t)$ is the sinc function of \eqref{phi}, we further reduce the error of $P_\calC u(t)$ with $x(t)$ (the subscript $\scriptscriptstyle\mathrm{P}$ in $M\subP$ has been chosen in reference to the POCS method that will be introduced later on). We thus obtain the error reductions
\begin{equation}\label{err-decr}
\|M\subP u-x\|\leq\|P_\calC u-x\|\leq\|u-x\|
\end{equation}
where at least one of the two inequalities is strict when $u(t)\notin\calB\cap\calC$, i.e., when $u(t)$ is not simultaneously bandlimited and consistent (note that $u(t)\in\calC\backslash\calB$ implies that $P_\calC u(t)=u(t)\notin\calB$). For repetitive improvements, this suggests the use of the algorithm
\begin{equation}\label{u-rec}
u\up{k+1}(t):=M\subP\,u\up{k}(t),\qquad k\geq0.
\end{equation}
As a common mathematical notation, one writes
$$u\up{k}(t)=M\subP^k\,u\up{0}(t),\qquad k\geq0.$$
From the set theoretic viewpoint, $\varphi(t)*v(t)$ is nothing but the orthogonal projection of $v(t)$ onto $\calB$, which is a closed linear subspace of $L^2(D)$. Thus,
$$M\subP=P_\calB P_\calC.$$
Thus, the estimates $u\up{k}(t)$ are obtained by alternating orthogonal projections between $\calC$ and $\calB$. This is a particular case of the
method of projection onto convex sets (POCS) \cite{Combettes93,Bauschke96}.
We know that $\|u\up{k}-x\|$ is strictly decreasing with $k$ as long as $u\up{k}(t)$ is not in $\calB\cap\calC$. In fact, it is known that $u\up{k}(t)$ must eventually converge to an element of $\calB\cap\calC$.  Moreover, as $\calB$ and $\calC$ are convex sets that are more  specifically affine spaces,
\begin{equation}\label{u-lim}
u\up{\infty}(t):=\lim_{k\rightarrow\infty}M\subP^k u\up{0}(t)=P_{\calB\cap\calC}\,u\up{0}(t)
\end{equation}
where the limit is in the sense of $L^2(D)$. Thus, $u\up{\infty}(t)$ is the element of $\calB\cap\calC$ that is closest to the initial estimate $u\up{0}(t)$ in the MSE sense. Finally, it follows from \eqref{PCu-explicit} and \eqref{bandlimit-proj} that
\begin{equation}\label{PBPC}
M\subP u(t)=u(t)+\smallsum{n\in\N}(\theta_n-\langle h_n,u\rangle)\,\midfrac{\widetilde h_n(t)}{\|h_n\|^2}
\end{equation}
for all $u(t)\in\calB$.

\subsection{Numerical experiments}\label{subsec:POCS-exp}

\begin{figure}
\centerline{\hbox{\scalebox{0.61}{\includegraphics{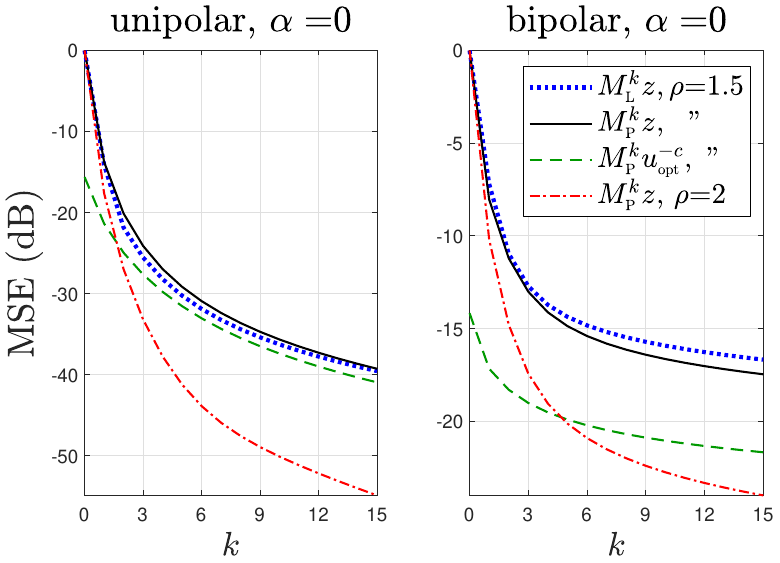}}}}
\vspace{3mm}
\centerline{\hbox{\scalebox{0.61}{\includegraphics{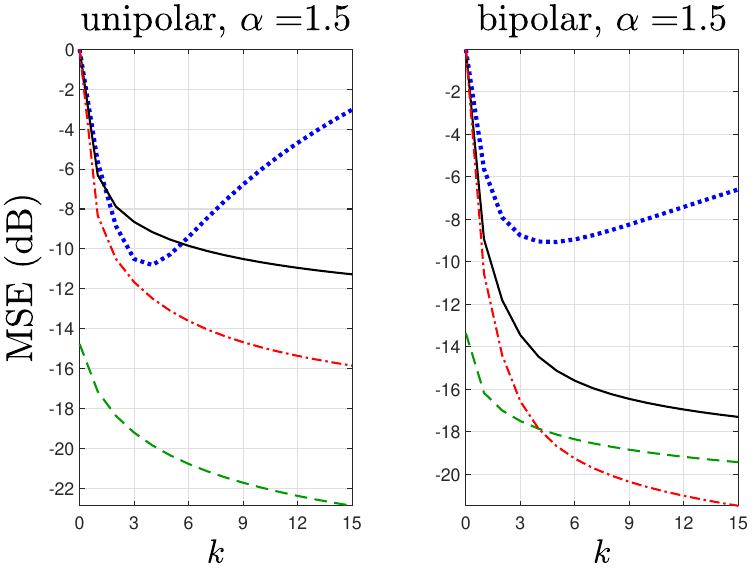}}}}
\vspace{3mm}
\centerline{\hbox{\scalebox{0.61}{\includegraphics{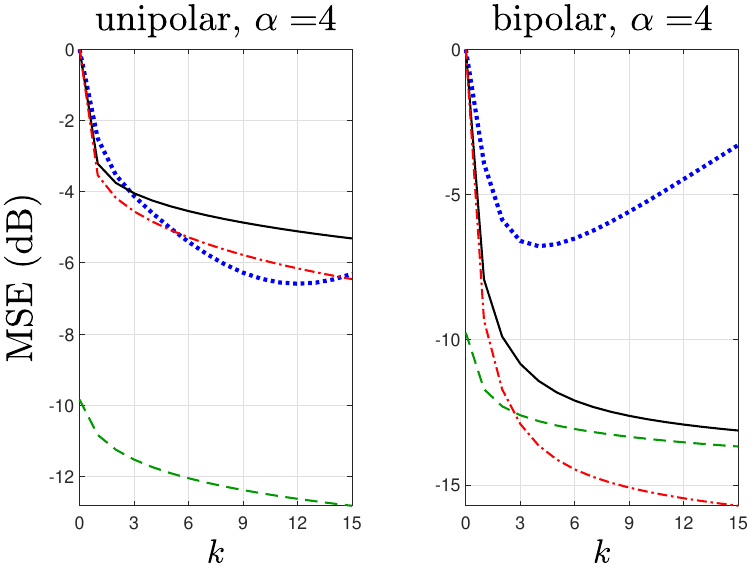}}}}
\caption{MSE performance of the iterates $M\subL^k z(t)$, $M\subP^k z(t)$, $M\subP^k u\subsmall{\mathrm{opt}}^{-c}(t)$ at the oversampling ratio $\rho=1.5$, as well as $M\subP^k z(t)$ at the oversampling ratio $\rho=2$.}\label{figiter}
\vspace{-2mm}
\end{figure}
\begin{table}
\renewcommand{\arraystretch}{1.5}
\centerline{\begin{tabular}{|c|c||c|l|c|l|}
\cline{4-6}
\multicolumn{3}{c|}{} & \multicolumn{1}{c|}{(a)} & (b) & \multicolumn{1}{c|}{(c)}\\
\cline{4-6}
\multicolumn{3}{c|}{} & \multicolumn{3}{c|}{$I-RS$}\\
\cline{4-6}
\multicolumn{3}{c|}{} & \multicolumn{2}{c|}{\raisebox{-0.5ex}{operator norm}} & \multicolumn{1}{c|}{\raisebox{-0.5ex}{$^{\textstyle\text{spectral}^{}}_{\textstyle\text{ radius}_{}}$}}\\[1ex]
\cline{2-6}
\multicolumn{1}{c|}{} & $\alpha$ & $\Delta\subsmall{\max}$ & \multicolumn{1}{c|}{POCS} & \multicolumn{2}{c|}{Lazar's method} \\
\hline
\hline
& $0$ &$1.8$ &$0.79$ & $0.77$ & $0.76$ \\
\cline{2-6}
unipolar & $1.5$ & $4.3$ & $1-3{\cdot}10^{-3}$ & $1.19$ & $1.0002$ \\
\cline{2-6}
& $4$ &$5.5$ & $1-2{\cdot}10^{-6}$ & $1.12$ & $1.03$ \\
\hline
& $0$ & $2.8$ & $1-4{\cdot}10^{-3}$ & $1.76$ & $1-4{\cdot}10^{-3}$ \\
\cline{2-6}
bipolar & $1.5$ & $3.2$ & $1-2{\cdot}10^{-4}$ & $1.25$ & $1.008$ \\
\cline{2-6}
& $4$ &$5.1$ & $1-8{\cdot}10^{-7}$ & $1.13$ & $1.03$\\
\hline
\end{tabular}}
\pp\pp
\caption{}\label{tab:coefs}
\end{table}
Under the same experimental conditions as in Fig. \ref{fig1}, we compare in Fig. \ref{figiter} the MSE of estimates from various iterative methods starting from one of the following two initial estimates:
$$z(t):=0\qquad\mbox{or}\qquad u\subsmall{\mathrm{opt}}^{-c}(t):=u\subsmall{\mathrm{opt}}(t)-c$$
remembering that $u\subsmall{\mathrm{opt}}(t)$ is an estimate of $x^c(t)=x(t)+c$. The tested iterates are specifically
$$M\subL^k z(t),\qquad M\subP^k z(t),\qquad M\subP^k u\subsmall{\mathrm{opt}}^{-c}(t)$$
where $M\subL$ is the mapping of the first iterative algorithm of bandlimited reconstruction for LIF proposed by Lazar in \cite{LAZAR2005401} (the subscript $\scriptscriptstyle\mathrm{L}$ in $M\subL$ has been chosen in reference to the name of Lazar).
The mapping $M\subL$ is explicitly defined by
\begin{eqnarray}\label{Lazar-iter}
M\subL u(t):=u(t)+\smallsum{n\in\N}
\big(\theta_n{-}\langle h_n,u\rangle\big)\,\varphi(t{-}\tau_n)
\end{eqnarray}
where $\tau_n:=(t_{n-1}{+}t_n)/2$ and $\varphi(t)$ is defined in \eqref{phi}. Further details on Lazar's method will be given in Section \ref{sec:contract}. These iterates are all tested on the same LIF outputs with an average density of firing instants of $\rho=1.5$ with respect to the Nyquist period. The iterates $M\subP^k z(t)$ are additionally tested at the oversampling ratio $\rho=2$. Our observations are as follows.
\begin{enumerate}
\item $M\subL^kz(t)$ is outperformed by $M\subP^k z(t)$ at $\rho=1.5$, except in the leakage-free unipolar case where it yields a similar MSE. Moreover, $M\subL^kz(t)$ diverges with $k$ in the leaky cases. These are evidently cases where the sufficient condition for convergence derived in \cite{LAZAR2005401} is not met in spite of the oversampling.
\item By applying $M\subP^k$ to $u\subsmall{\mathrm{opt}}^{-c}(t)$, one sees an example where the POCS method can be used to improve an estimate that is provided by another method. At the same time, one obtains a head start in MSE reduction compared to $M\subP^k z(t)$, which is particularly substantial in the leaky unipolar cases. This however requires the availability of input-output encoding statistics.
\item From the iterates $M\subP^k z(t)$, we see by how much the POCS convergence is accelerated by increasing the sampling density from 1.5 to 2. The acceleration is particularly dramatic with the leakage-free unipolar configuration, where the number of iterations necessary to reach approximately an MSE of -35dB is reduced by a factor of 3.
\item The rate of convergence of the POCS method is seen to slow down with the  increase of the biggest gap $\Delta\sublow{\max}:=\max\nN\Delta t_n$ between the firing instants, whose average over the trials is reported in Table \ref{tab:coefs}. This gap appears to increase with larger leakage, but also when going from the unipolar to the bipolar configuration in the case of zero leakage.
\end{enumerate}
More theoretical comparisons between the POCS and Lazar's methods are performed in Sections \ref{sec:contract} and \ref{sec:noise}.

\section{Nonuniform sampling pseudo-inversion}\label{sec:nonunif-samp}

Until now, we have mostly focused on the algorithmic aspects of the POCS method with iterative MSE decrease, guaranteed convergence and numerical observations. There are however a number of pending questions: in what fundamental way do the POCS method and Lazar's algorithm differ from each other? what is the meaning of the POCS iteration limit when reconstruction is not unique and/or when the sampling is corrupted by noise? There is in fact a whole theoretical background of linear algebra in Hilbert spaces to be used to address these questions. The sampling data of \eqref{equ-inf3} theoretically amounts to providing the image of $x(t)$ through a linear operator $S$ from $\calB$ to $\RR^N$. We show in this section that the POCS iteration converges to a certain pseudo-inverse of $S$, which will help answer the above question in the present and the next two sections.

But another motivation is to bring new insight to the problem of signal reconstruction in LIF encoding by placing it in a more general and abstract context of generalized nonuniform sampling. This allows a more fundamental and unified analysis of this problem which simultaneously connects it with existing research from the past \cite{Yen56} as well as points the potential of the present techniques for future generalizations.

\subsection{Generalized sampling}

We saw that the explicit information the IFS encoder provides about its input $x(t)$ lies in the inner-product values of \eqref{equ-inf3}. For a generalized study of signal reconstruction, we can present \eqref{equ-inf3} in the form
\begin{equation}\label{equ-inf-gen}
\forall n\in\N,\qquad \langle s_n,x\rangle=\theta_n
\end{equation}
where $s_n(t)\in\calB$ for all $n\in\N$. Indeed
\begin{equation}\label{hth}
\forall u(t)\in\calB,\qquad\langle h_n,u\rangle=\langle\widetilde h_n,u\rangle
\end{equation}
since $\langle h_n-\widetilde h_n,u\rangle=0$ by orthogonality in the frequency domain. Thus \eqref{equ-inf3} is equivalent to \eqref{equ-inf-gen} with $s_n(t):=\widetilde h_n(t)\in\calB$ for all $n\in\N$.
But for general analysis, we will only assume that
$(s_n(t))\nN$ is some known sequence of functions of $\calB$. The presentation of \eqref{equ-inf-gen} as generalized samples of $x(t)$ was previously introduced in \cite{eldar2005general}. Depending on the properties of $(s_n(t))\nN$, we will have various conclusions on the potential signal reconstructions.

\subsection{Sampling operator and inversion}\label{subsec:sampop}

Consider the linear operator
\begin{equation}
S:\begin{array}[t]{ccl}
\calB & \rightarrow & \RR^N\\
u(t) & \mapsto & \big(\langle s_n,u\rangle\big)\nN
\end{array}.\label{S}
\end{equation}
This is usually called an {\em analysis operator} in frame theory \cite{Christensen08}. We will specifically call it a {\em sampling operator} in this paper. Using the vector notation $\btheta=(\theta_n)\nN$, \eqref{equ-inf-gen} tells us that
\begin{equation}\label{equ-op}
S x(t)=\btheta.
\end{equation}
To retrieve $x(t)$, one then basically needs to ``invert'' $S$. However, $S$ is not invertible when reconstruction is not unique. All that can be rigorously defined is the set of consistent estimates
\begin{equation}\label{inv-set}
S^{-1}(\btheta):=\big\{u(t)\in\calB:S u=\btheta\big\}.
\end{equation}
The first step is to characterize this set for any given $\btheta\in\RR^N$. Clearly,
$$S^{-1}(\btheta)\neq\emptyset\qquad\Leftrightarrow\qquad\btheta\in\ran(S)$$
where $\ran(S)$ denotes the range of $S$. This is automatically the case when $\btheta$ is  precisely defined by \eqref{equ-op}. Whenever non-empty, $S^{-1}(\btheta)$ has some basic characterization from linear algebra (see \cite[\S7.2]{Garcia17} and \cite[\S2.5]{Christensen08} for various contexts), which we provide and prove in the next proposition for our specific assumptions and notation. This is based on the linear span $\scS$ of $(s_n(t))\nN$,
$$\scS:=\mathrm{span}(s_n(t))\nN.$$

\begin{proposition}\label{prop:sol-set}
For any $\btheta\in\ran(S)$,
\begin{enumerate}[label=(\roman*)]
\item $S^{-1}(\btheta)$ contains a unique solution $x_\btheta(t)$ in $\scS$,\vspace{0.5mm}
\item
\begin{equation}\label{sol-set}
S^{-1}(\btheta)=x_\btheta(t)+\scS^\perp
\end{equation}
where $\scS^\perp$ is the orthogonal complement of $\scS$ in $\calB$,\vspace{0.5mm}
\item $x_\btheta(t)$ is the minimum-norm element of $S^{-1}(\btheta)$.
\end{enumerate}
\end{proposition}
\line
\begin{IEEEproof}
For any $u\in\calB$ and $v\in S^{-1}(\btheta)$,
\begin{align}
u\in S^{-1}(\btheta)\quad\Leftrightarrow\quad S(u{-}v)=Su-Sv=\btheta-\btheta=0\nonumber\\
\Leftrightarrow\quad\langle s_n,u{-}v\rangle=0,\forall n\in\N\quad\Leftrightarrow\quad u-v\in\scS^\perp.\label{equivs}
\end{align}
As by assumption $S^{-1}(\btheta)\neq\emptyset$, let $x_\btheta:=P_\scS v_0\in\scS$ for an arbitrary $v_0\in S^{-1}(\btheta)$. Since $x_\btheta-v_0\in\scS^\perp$, \eqref{equivs} implies that $x_\btheta\in S^{-1}(\btheta)$. For any $u\in\scS\cap S^{-1}(\btheta)$, \eqref{equivs} implies that $u-v_0\in\scS^\perp$ and hence $u=P_\scS v_0=x_\btheta$. This proves (i). Meanwhile, (ii) follows from \eqref{equivs} with $v=x_\btheta\in\scS$. As $u=x_\btheta+(u-x_\btheta)$ where $x_\btheta\in\scS$ and $u-x_\btheta\in\scS^\perp$ for any $u\in S^{-1}(\btheta)$, it follows from the Pythagorean theorem that $\|u\|^2\geq\|x_\btheta\|^2$. This proves (iii).
\end{IEEEproof}

\subsection{Sampling pseudo-inverse}\label{subsec:samp-pseudo}

In absence of an inverse for $S$, a classic approach to the estimation $x(t)$ is to consider the Moore-Penrose pseudo-inverse $S^\dagger$ of $S$ \cite[\S6.11]{Luenberger69}. For any $\btheta\in\RR^N$,
\begin{eqnarray}
&\displaystyle S^\dagger\btheta:=\argmin_{u(t)\in\calD(\btheta)}\|u\|\quad\label{pseudo0}\\
\mbox{where}&\displaystyle\calD(\btheta):=\Big\{u(t)\in\calB: \|Su-\btheta\| \mbox{ is minimized}\Big\}&\quad~\nonumber
\end{eqnarray}
with a norm $\|\cdot\|$ in $\RR^N$ that is induced by some chosen inner-product $\langle\cdot,\cdot\rangle$ of $\RR^N$ (while using the same notation of norm and inner-product as in $\calB$, it is the argument that removes any ambiguity).
Since $\|\cdot\|$ in $\RR^N$ is induced by $\langle\cdot,\cdot\rangle$, $\|Su-\btheta\|$ is minimized for a given $\btheta\in\RR^N$ if and only if $Su$ is equal to the orthogonal projection $P_{\ran(S)}\btheta$ of $\btheta$ onto $\ran(S)$ in the sense of $\langle\cdot,\cdot\rangle$. Thus, we also have
\begin{equation}\label{D2}
\calD(\btheta)=S^{-1}(\bbtheta)\qquad\mbox{where}\qquad\bbtheta:=P_{\ran(S)}\btheta.
\end{equation}
Since $\bbtheta\in\ran(S)$, it follows from Proposition \ref{prop:sol-set}\,(iii) that
\begin{equation}\label{pseudo2}
S^\dagger\btheta=x_\bbtheta(t).
\end{equation}
The default approach is to take $\|\cdot\|$ in $\RR^N$ equal to the Euclidean norm ${\|\cdot\|_2}$. In this case, we specifically denote the pseudo-inverse by $S_2^\dagger$. Otherwise, when $\|\cdot\|$ is induced by some arbitrary inner-product $\langle\cdot,\cdot\rangle$ of $\RR^N$ , it can be shown that there exists an $N{\times}N$ positive definite matrix $W$ such that $\|\btheta\|=\|W\btheta\|_2$ for all $\btheta\in\RR^N$. In this situation, $S^\dagger$ can be seen as an extension of the definition of {\em weighted} pseudo-inverse introduced in \cite{elden1982weighted} for matrices.

\subsection{Limit of POCS iteration}

In Section \ref{subsec:POCS}, we introduced the POCS method to reconstruct $x(t)$ from its sample vector $Sx(t)=\btheta$ where the kernel functions of $S$ are
\begin{equation}\label{sh}
s_n(t):=\widetilde h_n(t),\qquad n\in\N
\end{equation}
and $(h_n(t))\nN$ can be {\em any} orthogonal family of $L^2(\RR)$ (see Proposition \ref{prop:PC}).
LIF encoding is the particular case where $h_n(t)$ is defined by \eqref{hn}, which we will not necessarily assume here. In practice, the orthogonality of the functions $(h_n(t))\nN$ is achieved as soon as their time supports do not overlap. This allows for example to replace the exponential factor in \eqref{hn} by any function of time. An estimate of $x(t)$ was obtained in \eqref{u-lim} by infinite iteration of the mapping $M\subP$ defined in \eqref{PBPC}. The goal here is to show that the limit of this iteration starting from a zero initial estimate leads to a weighted pseudo-inverse of $S$.
This result will be needed in particular when analyzing the POCS method under sampling noise in Section \ref{sec:noise}.
\line
\begin{proposition}\label{prop:POCS-lim}
Let $\btheta=(\theta_n)\nN$ be {\em any} element of $\RR^N$, $u\up{0}(t)$ be some initial estimate in $\scS$, and $u\up{k}(t):=M\subP^k u\up{0}(t)$ where $M\subP$ is defined by \eqref{PBPC} for the considered $(\theta_n)\nN$. Then,
\begin{equation}\label{u-lim-pseudo}
u\up{\infty}(t)=S\subP^\dagger\btheta
\end{equation}
where $S\subP^\dagger$ is the pseudo-inverse of $S$ with respect to the norm $\|\cdot\|\subP$ in $\RR^N$ induced by the weighted inner-product
\begin{equation}\label{weighted-inprod}
\langle\vc,\vc'\rangle\subP:=\smallsum{n\in\N}\midfrac{c_n c'_n}{\|h_n\|^2},\qquad\vc,\vc'\in\RR^N.
\end{equation}
\end{proposition}
\line
\begin{IEEEproof}
$S\subP^\dagger$ satisfies the properties of Section \ref{subsec:samp-pseudo} with $\|\cdot\|=\|\cdot\|\subP$ and $\langle\cdot,\cdot\rangle=\langle\cdot,\cdot\rangle\subP$, which we assume here.
For any given $\btheta\in\RR^N$, we denote by $M\subP^\btheta$ the mapping $M\subP$ defined in \eqref{PBPC} to highlight its dependence with $(\theta_n)\nN$. So explicitly, $u\up{\infty}=\lim_{k\rightarrow\infty}(M\subP^\btheta)^k u\up{0}$.

Assume that $\btheta\in\ran(S)$. It follows from \eqref{D2} and Proposition \ref{prop:sol-set}\,(ii) \eqref{C} that $$\calD(\btheta)=S^{-1}(\btheta)=x_\btheta+\scS^\perp=\calB\cap\calC$$
where the last equality is obtained by comparing \eqref{C} and \eqref{inv-set} based on \eqref{S}, \eqref{sh} and \eqref{hth}. So, on the one hand, $S\subP^\dagger\btheta=x_\btheta$ by Proposition \ref{prop:sol-set}\,(i). On the other hand, $x_\btheta=P_{\calB\cap\calC}u\up{0}$ since $u\up{0}\!-x_\btheta\in\scS$ and $\calB\cap\calC=x_\btheta+\scS^\perp$. One finally concludes that $S\subP^\dagger\btheta=u\up{\infty}$ from \eqref{u-lim}.

Consider now any $\btheta\in\RR^N$. Let $\bbtheta:=P_{\ran(S)}\btheta$. Since $P_{\ran(S)}\bbtheta=\bbtheta$, it is clear from \eqref{pseudo2} that $S\subP^\dagger\bbtheta=S\subP^\dagger\btheta$. Let $Q^\btheta:=M\subP^\bbtheta-M\subP^\btheta$. It is easy to derive from \eqref{PBPC} that
$Q^\btheta u=\sum\nN(\bar\theta_n\!-\theta_n)\widetilde h_n/\|h_n\|^2$. We deduce that
$$\langle Q^\btheta u,v\rangle=\textstyle\sum\nN(\bar\theta_n\!-\theta_n)\langle\widetilde h_n,v\rangle/\|h_n\|^2=\langle\bbtheta{-}\btheta,Sv\rangle\subP=0$$
for all $u,v\in\calB$, since $\bbtheta{-}\btheta\perp\ran(S)$. Thus, $M\subP^\btheta=M\subP^\bbtheta$. Then, $u\up{\infty}=\lim_{k\rightarrow\infty}(M\subP^\bbtheta)^k u\up{0}=S\subP^\dagger\bbtheta=S\subP^\dagger\btheta$ where the second equality results from the fact that $\bbtheta\in\ran(S)$.
\end{IEEEproof}

\subsection{Sampling situations}

The power of pseudo-inversion is that it defines an inversion procedure of $S$ in all possible sampling situations. We show in Table \ref{tab} these various situations. We say that reconstruction is unique when $S^{-1}(\btheta)$ is a singleton (i.e., contains a unique element) for any $\btheta\in\ran(S)$. The main structure of Table \ref{tab} is justified by the following proposition.
\line
\begin{proposition}\label{prop:unique-rec}
The following statements are equivalent:
\begin{enumerate}[label=(\roman*)]
\item $S^{-1}(\btheta)$ is a singleton for some $\btheta\in\ran(S)$,\vspace{0.5mm}
\item $S^{-1}(\btheta)$ is a singleton for all $\btheta\in\ran(S)$,\vspace{0.5mm}
\item $S$ is injective, i.e., $S u=0$ only when $u=0$,\vspace{0.5mm}
\item $\scS=\calB$.
\end{enumerate}
\end{proposition}
\line
\begin{IEEEproof}
(iv) is equivalent to $\scS^\perp=\{0\}$. One then easily sees from \eqref{sol-set} the equivalences between (iv), (i) and (ii). The equivalence between (ii) and (iii) is a basic result of linear algebra.
\end{IEEEproof}
\ppnoi
In Table \ref{tab} we have omitted the case where
$\scS$ and $\ran(S)$ are simultaneously proper subspaces of $\calB$ and $\RR^N$, respectively,
which is unlikely to happen in data acquisition. In absence of this pathological case, $N\geq\dim\calB$ is a necessary and sufficient condition for uniqueness of reconstruction. This is the case where $D=[0,T]$ with $T\leq N$, since $\dim(\calB)=T$ given our setting of Nyquist period 1 in Section \ref{subsec:sig-setting}. When $N<\dim\calB$, we say that the sampling is incomplete. The term of ``uniqueness of reconstruction'' specifically applies to the situation of noise-free sampling. Noise-corrupted sampling will be studied in Section \ref{sec:noise}.

\begin{table}
\small
\renewcommand{\arraystretch}{1.4}
\centerline{\begin{tabular}{|c|c|c|c|}
\cline{2-4}
\multicolumn{1}{c|}{}&(a)&(b)&(c)\\
\cline{2-4}
\multicolumn{1}{c|}{} &  $\scS\varsubsetneq\calB$ & \multicolumn{2}{c|}{$\scS=\calB$\quad\quad} \\
\cline{2-4}
\multicolumn{1}{c|}{} & \multicolumn{2}{c|}{$\ran(S)=\RR^N$} & $\ran(S)\varsubsetneq\RR^N$ \\
\hline
\hline
\!$(s_n(t))\nN$ in $\calB$\! & independent & basis & \!overcomplete\!\\
\hline
$S$ & surjective& bijective &injective\\
\hline
$\rho:=N/\dim\calB$ & $<1$ & $=1$ & $>1$\\
\hline
sampling & incomplete & critical & oversampling\\
\hline
uniqueness of&&\multicolumn{2}{c|}{}\\[-1ex]
reconstruction & \raisebox{2mm}{no} & \multicolumn{2}{c|}{\raisebox{2mm}{yes\quad}}\\
\hline
sampling-noise&\multicolumn{2}{c|}{}&\\[-1ex]
filtering & \multicolumn{2}{c|}{\raisebox{2mm}{not possible}} & \raisebox{2mm}{possible}\\
\hline
\end{tabular}}
\pp\pp
\caption{}\label{tab}
\end{table}

\subsection{Early use of pseudo-inverse in nonuniform sampling}

The problem of reconstructing $x(t)$ from samples $Sx(t)=\btheta$ with $S$ of the type \eqref{S} was first considered in Yen's pioneering paper \cite{Yen56} for the point-sampling operator
\begin{equation}\label{point-S}
S:\begin{array}[t]{ccl}
\calB & \rightarrow & \RR^N\\
u(t) & \mapsto & \big(u(t_n)\big)\nN
\end{array}.
\end{equation}
This indeed takes the form of \eqref{S} with
$$s_n(t):=\varphi(t-t_n),\qquad n\in\N$$
where $\varphi(t)$ in the sinc function defined in \eqref{phi}, since
\begin{equation}\label{point-sampling}
\forall u(t)\in\calB,\qquad\langle s_n,u\rangle=(\varphi*u)(t_n)=u(t_n).
\end{equation}
In the context of bandlimited functions in $L^2(\RR)$ assumed in \cite{Yen56}, the functions $(s_n(t))\nN$ are independent but not complete in $\calB$. This falls in the case (a) of incomplete sampling in Table \ref{tab}, where reconstruction is not unique. Yen focused on the minimum-norm reconstruction. By means of Lagrange multipliers, the solution of this reconstruction was found in \cite[eq.(10)]{Yen56} to be
$$\hat x(t):=\smallsum{n\in\N}c_n\,\varphi(t-t_n)\qquad\mbox{where}\qquad
\vc:=\Phi^{-1}\btheta$$
and $\Phi$ is the $N{\times}N$ matrix of coefficients $\varphi(t_n\!-t_m)$ for $n,m\in\N$. It can be seen that
$$\hat x(t)=S^\dagger\btheta$$
as follows. After observing that $\varphi(t_n\!-t_m)=s_m(t_n)=\langle s_n,s_m\rangle$ due to \eqref{point-sampling}, one can see that $\Phi=SS^*$ where $S^*$ is adjoint of $S$ (see equ. (1.4) and (1.5) of \cite{Christensen08} with $S=T^*$, which implies that $S^*=T$). One eventually finds that
$\hat x(t)=S^*(SS^*)^{-1}\btheta$. It then follows from \cite[equ.(2.11)]{Christensen08} that $\hat x(t)=S^\dagger\btheta$ given that $\ran(S)=\RR^N$ since $(s_n(t))\nN$ are independent.
Note that the norm $\|\cdot\|$ in $\RR^N$ for which $S^\dagger$ is defined here need not be specified since $\calD(\btheta)$ in \eqref{D2} is systematically equal to $S^{-1}(\btheta)$.

\section{Framework of contraction algorithms}\label{sec:contract}

In Section \ref{subsec:POCS-exp}, we compared numerically the POCS method with the prior algorithm designed by Lazar in \cite{Lazar04,LAZAR2005401}. The purpose of this section is to analyze these two methods at a more theoretical level. We show in particular that they actually belong to a single class of algorithms based on contraction mappings. This gives additional tools to understand the difference between these two methods in convergence properties.

\subsection{Class of algorithms}

As can be seen in \eqref{PBPC} and \eqref{Lazar-iter}, the POCS and Lazar's methods are both based on a recursion of the type
\begin{eqnarray}
&u\up{k+1}(t)=Mu\up{k}(t)\label{gen-iter}\\[-0.5ex]
\lefteqn{\hspace{-6mm}\mbox{where}}\nonumber\\[0ex]
&\hspace{-6mm}Mu(t):=u(t)+
\smallsum{n\in\N}\big(\theta_n{-}\langle s_n,u\rangle\big)\,r_n(t),\quad u(t)\in\calB.\label{M}
\end{eqnarray}
While the functions $(s_n(t))\nN$ are imposed by the encoder (given by \eqref{sh} and \eqref{hn} in LIF encoding), the functions $(r_n(t))\nN$ depend on the method. The intuition behind this iteration is as follows. For an early intuition about the design of this mapping, note the following.

\begin{remark}\label{rem0}
\begin{enumerate}[label=(\roman*)]
\item Every function $u(t)\in\scS^{-1}(\btheta)$ is a fixed point of $M$.
\item If the sequence $(u\up{k}(t))_{k\geq0}$ is convergent, then its limit $u\up{\infty}(t)$ is a fixed point of $M$.
\end{enumerate}
\end{remark}
\line  Evidently, (ii) does not imply that $u\up{\infty}(t)\in S^{-1}(\btheta)$. However, this can be ensured with some additional requirement on $M$. A typical condition that one tries to create is the following.
\line
\begin{condition}\label{cond}
\begin{enumerate}[label=(\roman*)]
\item $\mathrm{span}(r_n(t))\nN=\scS$,
\item $M$ is a {\em contraction} in $\scS$, i.e., there exists $\gamma\in[0,1)$ such that
\begin{equation}\label{contraction}
\forall u(t),v(t)\in\scS,\qquad\|Mu-Mv\|\leq\gamma\|u-v\|.
\end{equation}
\end{enumerate}
\end{condition}
Note that (i) implies that $\scS$ is invariant under $M$ as can be seen in \eqref{M}, so that the contraction property is usable. The contraction mapping theorem \cite[\S1.2]{smart1980fixed} then implies that $M$ has a unique fixed point $f_M(t)$ in $\scS$.
\line
\begin{proposition}\label{prop:contract}
Let $\btheta=(\theta_n)\nN$ be any element $\RR^N$ and $M$ be defined by \eqref{PBPC} for the considered $(\theta_n)\nN$ while satisfying Condition \ref{cond}. Then
\begin{enumerate}[label=(\roman*)]
\item $M$ has a unique fixed point $f_M(t)$ in $\scS$.
\item For any initial estimate $u\up{0}(t)$ in $\scS$, the iterates $u\up{k}(t):=M^k u\up{0}(t)$ tend to
\begin{equation}\label{u-lim0}
u\up{\infty}(t)=f_M(t)
\end{equation}
with the following error decay
\begin{equation}\label{conv-rate}
\|u\up{\infty}-u\up{k}\|\leq \gamma^k\|u\up{\infty}-u\up{0}\|,\qquad k\geq0.
\end{equation}
\item If $\btheta\in\ran(S)$, then
$f_M(t)=x_\btheta(t)$.
\end{enumerate}
\end{proposition}
\line
\begin{IEEEproof}
(i) This is a result of the contraction mapping theorem \cite[\S1.2]{smart1980fixed} within the space $\scS$.

(ii) We have $\|f_M-u\up{k+1}\|=\|M f_M-Mu\up{k}\|\leq{\gamma\|f_M-u\up{k}\|}$ from \eqref{contraction}. This leads to $\|f_M-u\up{k}\|\leq \gamma^k\|f_M-u\up{0}\|$ by induction. This proves \eqref{u-lim0} as $\gamma\in[0,1)$, and hence \eqref{conv-rate}.

(iii) Given that $\btheta\in\ran(S)$, the function $x_\btheta$ defined in Proposition \ref{prop:sol-set} is in $\scS$ while being a fixed point of $M$ due to Remark \ref{rem0}\,(i). So $f_M=x_\btheta$.
\end{IEEEproof}
\ppnoi
The following gives as a consequence a sufficient condition for perfect reconstruction.
\line
\begin{proposition}\label{prop:perfect-rec}
Under the conditions of Proposition \ref{prop:contract}, assume moreover that $\scS=\calB$ and $\btheta=Sx(t)$. Then, regardless of the initial estimate $u\up{0}(t)\in\calB$,  $$u\up{\infty}(t)=x(t).$$
\end{proposition}

\begin{IEEEproof}
Since $\scS=\calB$, it follows from the statements (ii) and (iii) of Proposition \ref{prop:contract} that $u\up{\infty}(t)=f_M(t)=x_\btheta(t)$. But due to Proposition \ref{prop:unique-rec}, $S^{-1}(\btheta)$ must be a singleton. Since both $x_\btheta(t)$ and $x(t)$ are in $S^{-1}(\btheta)$, $x_\btheta(t)=x(t)$.
\end{IEEEproof}
\ppnoi
While Proposition \ref{prop:contract} gave standard results on contractions within $\scS$, the next proposition shows that Condition \ref{cond} ensures the convergence of the iterates $u\up{k}(t)$ of \eqref{gen-iter} for any initial estimate $u\up{0}(t)$ in $\calB$.
\line
\begin{proposition}\label{prop:gen-init}
Assume the conditions of Proposition \ref{prop:contract} with the difference that $u\up{0}(t)$ is {\em any} initial estimate in $\calB$. Then, the iterates $u\up{k}(t):=M^k u\up{0}(t)$ tend to
\begin{equation}\label{u-lim-gen}
u\up{\infty}(t)=f_M(t)+u\up{0}(t)-P_\scS u\up{0}(t)
\end{equation}
with the error decay of \eqref{conv-rate}.
\end{proposition}
\line
\begin{IEEEproof}
Let $v\up{k+1}:=Mv\up{k}$ for $k\geq0$ starting from $v\up{0}:=P_\scS u\up{0}\in\scS$. We know by Proposition \ref{prop:contract}\,(ii) that $v\up{\infty}=f_M$. Let us show that
\begin{equation}\label{uv-iter}
u\up{k}=v\up{k}+d\qquad\mbox{where}\quad
d:=u\up{0}\!-P_\scS u\up{0}.
\end{equation}
This is clearly true at $k=0$.
Because $d\in\scS^\perp$, then $\langle s_n,d\rangle=0$ for all $n\in\N$. It is then easy to see that
$M(v\up{k}\!+d)=M(v\up{k})+d=v\up{k+1}+d.$ This allows a recursive proof of \eqref{uv-iter} for $k\geq0$. This leads to \eqref{u-lim-gen}. As \eqref{conv-rate} is satisfied by $v\up{k}$, it is also satisfied by $u\up{k}$ by mere translation by $d$.
\end{IEEEproof}

\subsection{Linear operator analysis}

The next step is to find a more explicit expression of Condition \ref{cond}. For the chosen functions $(r_n(t))\nN$, consider the linear operator
\begin{align}
R&:\begin{array}[t]{ccl}
\RR^N & \rightarrow &\calB\\
(c_n)\nN & \mapsto & \!\!\smallsum{n\in\N}c_n\,r_n(t)
\end{array}\qquad\label{R}
\end{align}
which is usually called a {\em synthesis operator} in frame theory \cite{Christensen08}.
Using the vector notation $\btheta=(\theta_n)\nN$, $M$ takes the form
\begin{align}
Mu(t)&=u(t)+R\big(\btheta-S u(t)\big)\nonumber\\
&=(I-R S)u(t)+R\btheta\label{iter-op-gen}
\end{align}
where $I$ designates the identity operator on $\calB$. Thus, $M$ is an affine mapping of linear part $I-R S$. In this case, one simply obtains
$$Mu-Mv=(I-R S)(u-v).$$
Under Condition \ref{cond}\,(i), $I-RS$ leaves $\scS$ invariant. Then, \eqref{contraction} is realized with $\gamma:=\|I-R S\|_\scS$, where for any linear operator $A$ on a subspace $\scV$,
$$\|A\|_\scV:=\sup_{u\in\scV\backslash\{0\}}\midfrac{\|Au\|}{\|u\|}.$$
This value of $\gamma$ is more specifically the smallest possible value satisfying \eqref{contraction}. Thus, $M$ is a contraction in $\scS$ if and only if
$\|I-R S\|_\scS<1$. After noticing that $\mathrm{span}(r_n(t))\nN=\ran(R)$, then Condition \ref{cond} takes the following form.
\line
\begin{condition}\label{cond2}
$\ran(R)=\scS$ and $\|I-R S\|_\scS<1$.
\end{condition}

\subsection{POCS method}

We now return to the sampling kernel functions of \eqref{sh} which lead to the sampling operator
\begin{equation}\label{POCS-sampling}
Su(t):=\big(\langle\widetilde h_n,u\rangle\big)\nN,\qquad u(t)\in\calB
\end{equation}
where $(h_n(t))\nN$ is any orthogonal family of $L^2(\RR)$.
From \eqref{PBPC}, the mapping $M\subP$ of the POCS iteration is the particular case of $M$ in \eqref{M} where $r_n(t):=\widetilde h_n(t)/\|h_n\|^2$,
remembering again the identity \eqref{hth}.
With \eqref{iter-op-gen}, $M\subP$ then yields the compact expression
$$M\subP u(t)=(I-R\subP S)u(t)+R\subP\btheta,\qquad u(t)\in\calB$$
where
\begin{equation}\label{RP}
R\subP\vc:=\smallsum{n\in\N}c_n\,\widetilde h_n(t)/\|h_n\|^2,\qquad
\vc\in\RR^N.
\end{equation}
The following result was proved in \cite[\S IV.A]{Thao21a}\footnote
{The result of \cite{Thao21a} includes relaxation coefficients $\lambda_i$, which in the present case are all equal to 1.}.
\line
\begin{proposition}\label{prop:POCS-contract}
Let $S$ and $R\subP$ be defined by \eqref{POCS-sampling} and \eqref{RP} for any given orthogonal family $(h_n(t))\nN$ of $L^2(\RR)$. Then,
$$\|I-R\subP S\|_\scS<1.$$
\end{proposition}
\ppnoi
As $\ran(R\subP)=\mathrm{span}(\widetilde h_n(t))\nN=\scS$, then $S$ and $R\subP$ satisfy Condition \ref{cond2}. We thus retrieve with Proposition \ref{prop:contract} the fact from Proposition \ref{prop:POCS-lim} that $u\up{k}(t):=M\subP^k u\up{0}(t)$ is systematically convergent for any $u\up{0}(t)\in\scS$. By linking \eqref{u-lim0} and \eqref{u-lim-pseudo}, we find in particular that
\begin{equation}\label{MPSP}
u\up{\infty}(t)=f_{M\subP}(t)=S\subP^\dagger\btheta.
\end{equation}
We emphasize here that the convergence is independent of the conditions of sampling (see Table \ref{tab}).

In the experimental conditions of Fig. \ref{figiter} for the oversampling ratio $\rho=1.5$, we have reported the numerical values of $\|I-R\subP S\|_\scS$ in column (a) of Table \ref{tab:coefs}. Note here that $\scS=\calB$ as can be seen in Table \ref{tab} since $\rho>1$. More specifically, we have calculated the average $\overline{\gamma}$ of $\|I-R\subP S\|_\scS$ over the 1000 trial inputs for each sampling configuration. We can see that $\overline{\gamma}<1$ in all cases of column (a). We also observe that the closer $\overline{\gamma}$ is to 1, the slower the MSE decay is in Fig. \ref{figiter}.

\subsection{Lazar's method}\label{subsec:Lazar}

Lazar's method is also applicable to the sampling operator $S$ of \eqref{POCS-sampling} but with functions $(h_n(t))\nN$ that are specifically defined by \eqref{hn}. Under this restriction, the mapping $M\subL$ used by this algorithm and defined in \eqref{Lazar-iter} is the particular case of $M$ in \eqref{M} where $r_n(t):=\varphi(t{-}\tau_n)$ and $\varphi(t)$ is the sinc function defined in \eqref{phi}. With \eqref{iter-op-gen}, $M\subL$ then yields the expression
$$M\subL u(t)=(I-R\subL S)u(t)+R\subL\btheta,\qquad u(t)\in\calB$$
where
\begin{equation}\label{RL}
R\subL\vc:=\smallsum{n\in\N}c_n\,\varphi(t{-}\tau_n),\qquad\vc\in\RR^N.
\end{equation}
The leakage-free case ($\alpha=0$ in \eqref{hn}) was first analyzed in \cite{Lazar04} in the context of ASDM encoding. Based on mathematical results of \cite{Feichtinger94}, it was proved with $\N=\ZZ$ and $D=\RR$ that
\begin{equation}\label{Lazar-cond}
\|I-R\subL S\|_\calB\leq\Delta\subsmall{\max}:=\sup_{n\in\ZZ}\Delta t_n.
\end{equation}
Thus, by adjusting the IF encoding parameters so that $\Delta\subsmall{\max}<1$, then Condition \ref{cond2} is satisfied by $S$ and $R\subL$ with $\scS=\calB$. As a first remark, this algorithm is only applicable in the situation of unique reconstruction (cases (b) and (c) of Table \ref{tab} where $\scS=\calB$). The second remark is that $\Delta\subsmall{\max}<1$ is an even more restrictive condition. In finite dimension at least, only the average of $\Delta t_n$ needs to be less than 1 for unique reconstruction.

Now, $\Delta\subsmall{\max}<1$ is only a proved sufficient condition for Lazar's method to converge. Even though $\Delta\subsmall{\max}>1$ in all cases of Table \ref{tab}, we saw in Fig. \ref{figiter} that Lazar's iteration is still convergent when $\alpha=0$ in both unipolar and bipolar configurations. Using the same procedure as for the POCS method, we have reported in column (b)  of Table \ref{tab:coefs} the average $\overline{\gamma}$ of $\|I-R\subL S\|_\calB$. This value does appear to be less than 1 (and similar to the POCS value) in the leakage-free unipolar case. Surprisingly, $\overline{\gamma}$ is larger than $1$ in the leakage-free bipolar case. The convergence can however still be explained as the {\em spectral radius} of $I-R\subL S$ is less than 1 in this case (see \cite[p.506]{floudas2008encyclopedia}), as reported in column (c) of the table.

In the presence of leakage, an upper bound to $\|I-R S\|_\scS$ was later proposed in \cite{LAZAR2005401}. It was however expressed as an intricate function of the sequence $(\Delta t_n)_{n\in\ZZ}$ together with other parameters of the internal LIF encoder. Independently of the exact expression of this upper bound, Table \ref{tab:coefs}\,(b) shows numerically that $\|I-R\subL S\|_\scS>1$ in all leaky cases. Even the spectral radius of $I-R\subL S$ is larger than 1 in these cases as seen in  Table \ref{tab:coefs}\,(c).

\section{Unique reconstruction and noise behavior}\label{sec:noise}

In the previous section, we showed that the POCS and Lazar's methods consist in an iteration of the type
\begin{equation}\label{iter-op}
u\up{k+1}(t)=M u\up{k}(t):=(I-RS)u\up{k}(t)+R\btheta,\quad k\geq0.
\end{equation}
When $\scS=\calB$, $\btheta=Sx(t)$ and Condition \ref{cond2} is satisfied, we know from Proposition \ref{prop:perfect-rec} that $u\up{\infty}(t)=x(t)$, thus achieving perfect reconstruction. Note that this result is independent of the choice of synthesis operator $R$ as long as Condition \ref{cond2} is satisfied. In practice however, the sampling vector $\btheta$ is in general corrupted by noise,
\begin{equation}\label{noisy-samp}
\btheta=Sx(t)+\ve
\end{equation}
for some error vector $\ve$. By injecting this vector $\btheta$ into \eqref{iter-op}, one expects to obtain a deviated reconstruction $u\up{\infty}(t)=x(t)+e(t)$ for some error function $e(t)$. If $\ve\in\ran(S)$, no error reduction is possible since $\ve$ is undistinguishable from the samples of an actual bandlimited input component. This is always the case when $\ran(S)=\RR^N$ as reported in Table \ref{tab}\,(a-b). But when $\ran(S)\varsubsetneq\RR^N$, different operators $R$ may result in different deviations $e(t)$ with error vectors $\ve$ that are not in $\ran(S)$. We study this dependence in this section, and apply our analysis to the POCS and Lazar's methods with particular attention to time-quantization noise. As the condition $\ran(S)\varsubsetneq\RR^N$ corresponds to the case of Table \ref{tab}\,(c), we assume from now on that
\begin{equation}\nonumber\label{S=B}
\scS=\calB.
\end{equation}

\subsection{General limit of contraction algorithm}

Assuming Condition \ref{cond2} and $\scS=\calB$, we know from Proposition \ref{prop:contract}\,(ii) that, regardless of the initial estimate $u\up{0}(t)\in\calB$, $u\up{k}(t)$ converges to the unique fixed point  $f_M(t)$ of $M$ in $\calB$. Our present goal is to find an explicit expression of $u\up{\infty}(t)$ in terms of $S$, $R$ and $\btheta$. As a fixed point of $M$, it is easy to see from \eqref{iter-op-gen} that $u\up{\infty}(t)$ satisfies the equation
$$RS\,u\up{\infty}(t)=R\btheta.$$
But Condition \ref{cond2} with $\scS=\calB$ implies that $\|I-R S\|_\calB<1$. As a result of the Neumann series \cite[\S2.3]{kress2012linear}, it is then known that $RS$ is invertible.
Therefore,
\begin{eqnarray}
&u\up{\infty}(t)=\hat S\btheta\nonumber\\[0.5ex]
\mbox{where}\qquad\qquad&\hat S:=(R S)^{-1} R.&\qquad\qquad\qquad\label{hS}
\end{eqnarray}
Like $R$, $\hat S$ is a linear operator from $\RR^N$ to $\calB$, with however the specific property that
$$\hat S S=(R S)^{-1}R S=I.$$
This makes $\hat S$ a {\em left inverse} of $S$. This is precisely the required property for perfect reconstruction as
\begin{equation}\label{left-inv}
\btheta=Sx(t)\qquad\Rightarrow\qquad\hat S\btheta=\hat S S x(t)=x(t).
\end{equation}
This makes the restriction of $\hat S$ to $\ran(S)$ independent of $R$. However, due to the expression of $\hat S$ in \eqref{hS}, one will expect $\hat S\btheta$ to depend on $R$ when $\btheta\notin\ran(S)$.

\subsection{Left inversion and noise}

When $\btheta$ is given by \eqref{noisy-samp}, then
\begin{equation}\label{u-lim-noise}
u\up{\infty}(t)=\hat S\btheta=\hat S Sx(t)+\hat S\ve=x(t)+\hat S\ve.
\end{equation}
The question is which left inverse $\hat S$ is optimal for minimizing $\|\hat S\ve\|$.
A common approach for this question is to look at the pseudo-inverse $S^\dagger$ of $S$ defined in \eqref{pseudo0} for a given inner-product $\langle\cdot,\cdot\rangle$ in $\RR^N$.  The first observation is to see that $S^\dagger$ is a left inverse of $S$. Indeed, when $\btheta=S x$, it follows from \eqref{D2} that $\calD(\btheta)=S^{-1}(\btheta)$ which is reduced to $x(t)$ due to Proposition \ref{prop:unique-rec} given that $\scS=\calB$. So $S^\dagger S x=S^\dagger\btheta=x$. The second observation is the special action of $S^\dagger$ on a noise vector $\ve$. Consider the orthogonal decomposition
\begin{equation}\nonumber\label{e-decomp}
\ve=\ve\subin+\ve\subout\quad\mbox{where}\quad(\ve\subin,\ve\subout)\in\ran(S)\times\ran(S)^\perp
\end{equation}
where $\ran(S)^\perp$ is the orthogonal complement of $\ran(S)$ in $\RR^N$ with respect to $\langle\cdot,\cdot\rangle$. Since $\ve\subin\in\ran(S)$ and $S$ is injective, there exists a unique function $e\subin(t)\in\calB$ such that
$$S e\subin(t)=\ve\subin.$$
Although $\ve\subin$ is thought of a noise component, it is qualitatively undistinguishable from the samples of an actual bandlimited input.
Next, {\em regardless} of the left inverse $\hat S$, $\hat S\ve\subin=\hat S S e\subin(t)=e\subin(t)$. Thus,
\begin{equation}\label{noise-filtering}
\hat S\ve=e\subin(t)+\hat S\ve\subout.
\end{equation}
The magnitude of $\|\hat S\ve\subout\|$ depends on the choice of $\hat S$.
\line
\begin{proposition}\label{prop:pseudo-charac}
Let $\hat S$ be a left inverse of $S$. Then,
\begin{equation}\label{pseudo-charac}
\forall \ve\subout\in\ran(S)^\perp,~\hat S\ve\subout=0\qquad\Leftrightarrow\qquad\hat S=S^\dagger.
\end{equation}
\end{proposition}

\begin{IEEEproof}
$S^\dagger\btheta$ can be equivalently presented as the unique function $u(t)\in\calB$ such that $S u$ is the orthogonal projection $P_{\ran(S)}\btheta$ of $\btheta$ onto $\ran(S)$ with respect to $\langle\cdot,\cdot\rangle$, the uniqueness resulting from the injectivity of $S$. If $\ve\subout\in\ran(S)^\perp$, then $P_{\ran(S)}\ve\subout=0$, so $S^\dagger\ve\subout=0$. This proves the backward implication of \eqref{pseudo-charac}. For the forward implication, the left hand side of \eqref{pseudo-charac} implies that $\hat S$ coincides with $S^\dagger$ on $\ran(S)^\perp$. But $\hat S$ also coincides with $S^\dagger$ on $\ran(S)$ since  $(\hat S-S^\dagger)S=\hat SS-S^\dagger S=I-I=0$. So $\hat S=S^\dagger$.
\end{IEEEproof}
\ppnoi
In the traditional view of uniform sampling, $\ve\subin$ and $\ve\subout$ are nothing but the in-band and the out-of-band components of $\ve$, respectively. A left inverse $\hat S$ can be interpreted as a discrete-time lowpass filter followed by a sinc reconstruction that guarantees perfect recovery in absence of noise. Any out-of-band sample noise that is not completely eliminated results in aliasing. In our generalized framework of nonuniform sampling, the aliasing component is $\hat S\ve\subout$ in \eqref{noise-filtering}. It is completely eliminated with $\hat S=S^\dagger$.

\subsection{Application to LIF samplings}

Assuming the LIF sampling operator defined by \eqref{POCS-sampling} and \eqref{hn}, the left inverses obtained by the POCS and Lazar's methods are
\begin{equation}\nonumber
\hat S\subP:=(R\subP S)^{-1} R\subP\qquad\mbox{and}\qquad\hat S\subL:=(R\subL S)^{-1} R\subL
\end{equation}
where $R\subP$ and $R\subL$ are given in \eqref{RP} and \eqref{RL}, respectively. The goal is to compare $\hat S\subP$ and $\hat S\subL$ in terms of noise behavior. By connecting \eqref{u-lim-pseudo} and \eqref{MPSP} with $M=M\subP$, we actually have
$$\hat S\subP=S\subP^\dagger.$$
We recall that $S\subP^\dagger$ is the pseudo-inverse of $S$ with respect to the norm $\|\cdot\|\subP$ in $\RR^N$ induced by $\langle\cdot,\cdot\rangle\subP$ defined in \eqref{weighted-inprod}. Thus, the POCS method is better than Lazar's algorithm in eliminating the out-of-band noise components in the specific orthogonality sense of $\langle\cdot,\cdot\rangle\subP$. A central question is whether this inner-product is relevant. By default, one would tend to use the standard pseudo-inverse $S_2^\dagger$ relative to the canonical inner-product of $\RR^N$, which we introduced at the end of Section \ref{subsec:samp-pseudo}. We discuss this issue next.

\subsection{Qualitative analysis}

In this problem of sampling-noise reduction, there is a fundamental difference between uniform and nonuniform sampling. In the former case, there is a direct connection between the $\ell^2$-norm in the sample space (which is the Euclidean norm in finite dimension) and the $L^2$-norm of the continuous-time signals, via Parseval's identity. In this case, $S_2^\dagger$ is exactly the needed pseudo-inverse. In practice, the impact of noise has an equal weight in each sample with respect to reconstruction. But this ceased to be true when the sampling is nonuniform. In the case of \eqref{POCS-sampling}, it is intuitive that adding an error to a sample $\langle\widetilde h_n,x\rangle$ has less impact on the reconstruction of $x(t)$ when $\|\widetilde h_n\|$ is larger. It appears that the norm $\|\cdot\|\subP$ takes a better account of this biased sensitivity as can be seen in the expression of $\langle\cdot,\cdot\rangle\subP$ in \eqref{weighted-inprod}. The ultimate test will be to perform real numerical experiments on the left inverses $\hat S\subL$, $S\subP^\dagger$ and $S_2^\dagger$ under practical cases of sampling noise.

\subsection{Noise from time quantization}\label{subsec:time-quant}

The most intrinsic type of noise that time-encoding machines are subject to is from time quantization. The discretization of continuous values is inevitable in analog-to-digital conversion, except for the new situation here that it is performed in the time dimension instead of the traditional amplitude dimension. However, the resulting sampling errors can still be presented in the form of amplitude errors as implied by \eqref{noisy-samp}. The key here is to {\em systematically} define $S$ out of the quantized instants $(t_n)\nN$ output by the encoder, via the intermediate constructions of \eqref{POCS-sampling} and \eqref{hn}. Because these instants are deviated versions of the true instants $(t_n)\nN$ of \eqref{ineq-inf}, then the mathematical value of $Sx(t)$ becomes a deviated version of $\btheta$, which we can write in the form $\btheta-\ve$ where $\ve$ is unknown and viewed as noise. This leads to \eqref{noisy-samp}. While $S$ no longer reproduces the internal sampling mechanism of the encoder before quantization, we emphasize that it is deterministically constructed from the known output of the encoder. All errors in this approach lie in the vector $\ve$.

\subsection{Experimental results}

\begin{figure}
\centerline{\hbox{\scalebox{0.64}{\includegraphics{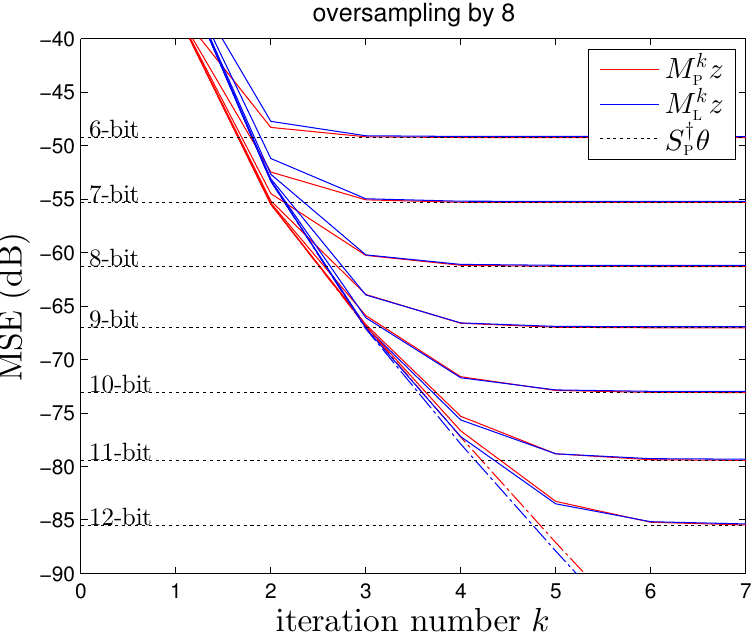}}}}
\caption{MSE of $M\subP^k z(t)$ and $M\subL^k z(t)$ versus the iteration number $k$, with the leakage-free unipolar configuration and various resolutions of time quantization (the dashed lines are without quantization). The MSE of $S\subP^\dagger\btheta$ is represented as a constant in dotted lines for each quantization resolution.}\label{fig:quant}
\vspace{2mm}
\centerline{\hbox{\scalebox{0.64}{\includegraphics{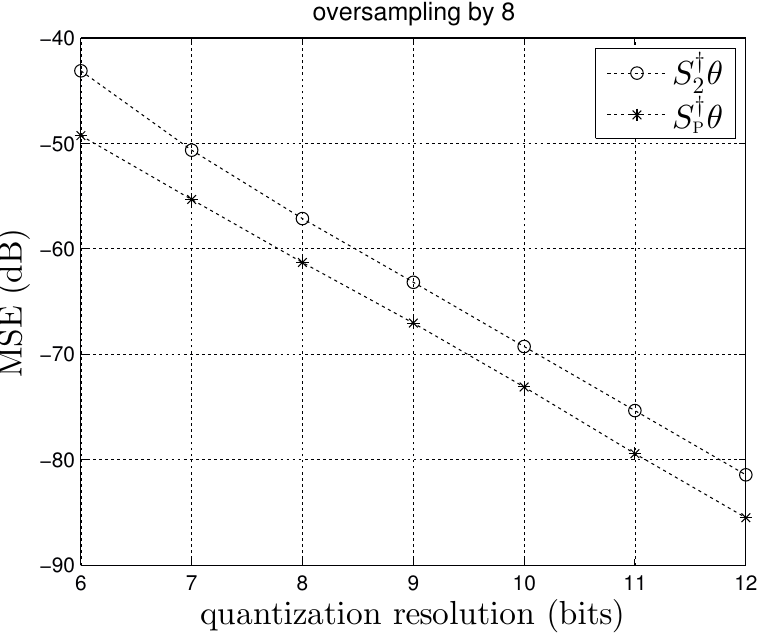}}}}
\caption{Reconstruction MSE of pseudo-inverses $S_2^\dagger\btheta$ and $S\subP^\dagger\btheta$ of $S$, with various resolutions of time quantization.}\label{fig:pseudo-quant}
\end{figure}

We show in Fig. \ref{fig:quant} the MSE of the iterates $M\subP^k z(t)$ and $M\subL^k z(t)$ resulting from time quantization with the leakage-free unipolar configuration of LIF in the same experimental conditions as in Fig. \ref{fig1}, with however an oversampling ratio of 8. We perform the experiments with various quantization resolutions of $(t_n)\nN$. We say that the resolution is $b$-bit when the quantization step size is the Nyquist period divided by $2^b$. The averaged values of $\|S\subP^\dagger\ve\|^2$ and $\|\hat S\subL\ve\|^2$ are extracted from the plot by looking at the MSE limits of $M\subP^k z(t)$ and $M\subL^k z(t)$ as $k$ increases, according to \eqref{u-lim-noise}. We also represent in dotted lines the MSE of $S\subP^\dagger\btheta$ which we calculate by direct computer implementation of $S\subP^\dagger$ for each bit resolution. The plots show that the MSE of both $M\subP^k z(t)$ and $M\subL^k z(t)$ tends to that of $S\subP^\dagger\btheta$. This is expected for $M\subP^k z(t)$ as a result of \eqref{u-lim-noise} with $\hat S:=\hat S\subP=S\subP^\dagger$. This however comes as an interesting result for $M\subL^k z(t)$, which shows that Lazar's method has a similar behavior to the POCS method towards time quantization. When zooming into the values of the plots, we actually observe that the POCS method outperforms Lazar's iteration by  approximately 0.1 dB in MSE decrease at every bit resolution.

But the more dramatic result is seen in Fig. \ref{fig:pseudo-quant} where, under the same experimental conditions, the MSE of $S\subP^\dagger\btheta$ is seen to be lower than that of $S_2^\dagger\btheta$ by 4 to 5 dB for all the tested bit resolutions. This tends to show the relevance of the weighted inner-product $\langle\cdot,\cdot\rangle\subP$ for capturing time-quantization noise. This does not imply that $S\subP^\dagger$ is the optimal left inverse for time-quantization noise. But this at least points that the standard pseudo-inverse $S_2^\dagger$ is not the appropriate left inverse for this type of encoding system.

\section{Discretization of the iterative algorithms}\label{sec:comp}

We recall from Section \ref{sec:contract} that the POCS and Lazar's methods consist in an iteration of the type
\begin{eqnarray}
&u\up{k+1}(t)=Mu\up{k}(t)\nonumber\label{gen-iter2}
\end{eqnarray}
where $M$ is given in \eqref{M}. The heavy part in the computation of $M$ is the continuous-time inner-product $\langle s_n,u\rangle$. As the argument $u(t)$ is bandlimited, this inner-product can be computed in discrete-time using the Nyquist-rate samples of $u(t)$. But this makes \eqref{M} a complicated operation as it requires a uniform sampling to obtain the coefficients of $(r_n(t))\nN$ in \eqref{M} which are themselves not uniformly distributed in time and not at the same rate as the Nyquist clock. Together with intricate algebra, this requires in practice the implementation of complicated buffering systems. There are however techniques to tackle these issues, which we present next.

\subsection{Zero initial estimate}\label{subsec:zero-case}

We recall from \eqref{iter-op} that $u\up{k}(t)$ satisfies a recursion of the type
\begin{equation}\label{u-opiter}
u\up{k+1}(t)=(I{-}R S)u\up{k}(t)+R\btheta.
\end{equation}
With $u\up{0}(t)=0$, one easily sees by induction that $u\up{k}(t)$ remains in $\ran(R)$ for all $k\geq0$. This implies the existence of a sequence $(\vc\up{k})_{k\geq0}$ of vectors in $\RR^N$ such that
\begin{equation}\nonumber
u\up{k}(t)=R\vc\up{k},\qquad k\geq0.
\end{equation}
Then, \eqref{u-opiter} is equivalent to
\begin{align}
R\vc\up{k+1}&=(I{-}R S)R\vc\up{k}+R\btheta\nonumber\\
\label{gen-iter3}
&=R\big((I{-}S R)\vc\up{k}+\btheta\big).
\end{align}
Inversely, consider the recursive iteration of the following system
\begin{subequations}\label{sys1}
\begin{align}
\vc\up{k+1}&:=(I{-}S R)\vc\up{k}+\btheta,\label{sys1a}\\
u\up{k}(t)&:=R\vc\up{k}\label{sys1b}
\end{align}
\end{subequations}
for $k\geq0$ starting from $\vc\up{0}=0$. Then, \eqref{sys1a} guarantees \eqref{gen-iter3}, which is equivalent to \eqref{u-opiter} based on \eqref{sys1b} with $u\up{0}(t)=0$. In practice, if one aims at the $q$th iterate $u\up{q}(t)$, one simply needs to iterate $q$ times the purely discrete-time transformation of \eqref{sys1a} starting from $\vc\up{0}=0$, then perform the discrete-time to continuous-time conversion of \eqref{sys1b} only {\em once} at $k=q$. Now, in \eqref{sys1a}, the mapping $SR$ is nothing but the linear operator on $\RR^N$ of matrix coefficients
\begin{equation}\label{SR}
SR=\Big[\langle s_m,r_n\rangle\Big]_{(m,n)\in\N^2}.
\end{equation}
One does not escape from having continuous-time inner-products as seen here. However, as these values do not depend on the iterates, they are to be calculated only once before the iteration.

\subsection{Precomputation of the matrix coefficients}

We derive the coefficients of \eqref{SR} for the application of LIF signal reconstruction by POCS. We recall in this case that $s_n(t):=\widetilde h_n(t)$ and $r_n(t):=\widetilde h_n(t)/\|h_n\|^2$, which yields
$$\langle s_m,r_n\rangle=\langle h_m,\widetilde h_n\rangle\,/\,\|h_n\|^2$$
remembering again the identity \eqref{hth}. Defining
\begin{equation}\label{Tmn}
T_n^m:=\int_{t_n}^{t_m}\dif t=t_m-t_n,
\end{equation}
we prove in Appendix \ref{app:innerprod} the following two results.
\line
\begin{proposition}\label{prop:thn1}
For any given function $f(t)$, let $f_n(t):=f(t-t_n)$. Then,
\begin{equation}\label{thn1}
\big\langle h_m, s_n\big\rangle=g(T^m_n)-e^{-\alpha\Delta t_m}g(T^{m-1}_n)
\end{equation}
where
\begin{equation}\label{g-thn1}
g(t):=\displaystyle\int_0^ t e^{\alpha(s-t)} f(s)\dif s.
\end{equation}
\end{proposition}

\begin{corollary}\label{prop:corol}
\begin{align}
\langle h_m,\widetilde h_n\rangle
&=e^{-\alpha\Delta t_n}\big(g(T^m_{n-1})-e^{-\alpha\Delta t_m}g(T^{m-1}_{n-1})\big)\nonumber\\
&\quad-\big(g(T^m_n)-e^{-\alpha\Delta t_m}g(T^{m-1}_n)\big)\label{inprod0}
\end{align}
where
\begin{align}
g(t)&:=\frac{1}{\alpha}\int_0^t\mathrm{sinh}(\alpha(t-s))\,\varphi( s)\,\dif  s.\label{g}
\end{align}
\end{corollary}
\ppnoi
The above function $g(t)$ only depends on the leakage coefficient $\alpha$. It can be precomputed and numerically stored in a lookup table. The computation of \eqref{inprod0} in terms of $(t_n)\nN$ then becomes simple, and only needs to be performed once before the iteration.
For the complete determination of $\langle s_m,r_n\rangle$, we need to derive $\|h_n\|^2$ that yields
\begin{equation}\label{nh2}
\|h_n\|^2=\int_{t_{n-1}}^{t_n}e^{-2\alpha(t_n-t)}\dif t
=\smallfrac{1}{2\alpha}\big(1-e^{-2\alpha\Delta t_n}\big).
\end{equation}

\subsection{General initial estimate}

There are several situations where it is more desirable to start the POCS iteration with a nonzero initial estimate $u\up{0}(t)$. Fig. \ref{figiter} showed reduced MSE results with $u\up{0}(t)=u\subsmall{\mathrm{opt}}^{-c}(t)$ for given iteration numbers $k$. When reconstruction is not unique, one may also wish to choose for $u\up{0}(t)$ some empirical guess of $x(t)$, as the limit $u\up{\infty}(t)$ is known from \eqref{u-lim} to be the element of $\calB\cap\calC$ that is closest to $u\up{0}(t)$. This technique of initial guess was previously used in \cite{Rzepka18} for POCS signal reconstruction from input level crossings.
The simple implementation of Section \ref{subsec:zero-case} can be used up to some space translation. This is specifically performed as follows.
\line
\begin{proposition}\label{prop:change-iter}
For any given $u\up{0}(t)\in\calB$, consider the functions $u\up{k}(t)$ recursively output by the system
\begin{subequations}\label{sys2}
\begin{align}
\vc\up{k+1}&:=(I{-}S R)\vc\up{k}+\bhtheta,\label{sys2a}\\
u\up{k}(t)&:=R\vc\up{k}+u\up{0}(t)\label{sys2b}
\end{align}
\end{subequations}
for $k\geq0$ starting from $\vc\up{0}=0$ with $\bhtheta:=\btheta-Su\up{0}$. Then $u\up{k}(t)$ satisfies \eqref{u-opiter} for all $k\geq0$.
\end{proposition}
\line
\begin{IEEEproof}
Let $v\up{k}(t):=u\up{k}(t)-u\up{0}(t)$. Then
\begin{align*}
(I{-}R S)v\up{k}+R\btheta=(I{-}R S)u\up{k}+R\btheta-(I{-}RS)u\up{0}\\
=u\up{k+1}-u\up{0}+RS u\up{0}=v\up{k+1}+RS u\up{0}.
\end{align*}
Thus, $v\up{k+1}=(I{-}R S)v\up{k}+R\bhtheta.$
From the results of Section \ref{subsec:zero-case}, we thus know that $v\up{k}=R\vc\up{k}$ where $\vc\up{k}$ is recursively defined by \eqref{sys2a} starting from $\vc\up{0}=0$ since $v\up{0}=0$. This leads to \eqref{sys2b}.
\end{IEEEproof}
\ppnoi
However, a price to pay for starting the iteration from any $u\up{0}(t)$ is the required derivation of $Su\up{0}=(\langle s_n,u\up{0}(t)\rangle)\nN$ for the construction of $\bhtheta$ involved in \eqref{sys2a}. Its computation complexity depends on $u\up{0}(t)$. If for example one chooses
$$u\up{0}(t)=\smallsum{n\in\N}\eps_n\,f(t-t_n)$$
for some function $f(t)$, like in \eqref{ulin}, then
$$\langle h_m,u\up{0}(t)\rangle=
\textstyle\sum\limits\nN\eps_n\,\langle h_m,f_n\rangle$$
where $f_n(t):=f(t-t_n)$. Then, the inner-products $\langle h_m,f_n\rangle$ are given in \eqref{thn1}.

\subsection{Future research}

An important aspect of the implementation is yet to be investigated. The matrix $SR$ involved in the discrete-time implementation of the POCS algorithm in \eqref{sys1a} of \eqref{sys2a} has a size $N$ that is virtually infinite compared to the practical windows of signal processing. A truncation of $SR$ automatically needs to be considered for the realistic implementation of these iterations, with expected performance degradations. Given the decay of the inner-products $\langle h_m,\widetilde h_n\rangle$ as $|m{-}n|$ increases, this is similar to the problem of FIR filter windowing. The difficulty here is that the filter is both time-varying and iterated. Some empirical truncation experiments have been performed in \cite{Thao21a} in the leakage-free case of signal reconstruction from ASDM outputs. But, beyond a needed generalization to leaky integration, this study deserves deeper theoretical investigations to be considered for future research.

\section{Conclusion}

While LIF encoding is commonly viewed as a neuron-inspired system that transforms continuous-time signals into spike trains, we approached the problem of bandlimited signal reconstruction from LIF outputs from the more general and abstract perspective of nonuniform sampling. In this view, each output spike gives the knowledge of an inner-product of the input with a kernel function that can be determined from the characteristics of the encoder. With the complete output of the LIF encoder, one thus has access to the transformation of the input by a known linear operator $S$, which we call the sampling operator. This more abstract picture allows to see better what exact information is available about the input in the LIF encoded output and outperform existing one-step input estimation techniques. We studied two iterative algorithms which mathematically achieve some type of inversion of $S$, Lazar's method \cite{Lazar04,LAZAR2005401} and the POCS method that were previously demonstrated on ASDM-based time encoding machines. While Lazar's method requires quite strict conditions to converge, such as a certain degree of oversampling and low leakage, the POCS method was shown to converge in all situations of sampling and leakage, including incomplete sampling. When the two methods converge simultaneously, they appear to have a similar behavior towards time-quantization noise, which is the most intrinsic type of noise in time encoding.

A simultaneous contribution of the paper was to perform this analysis at the most basic algebraic level of nonuniform sampling. This was the opportunity to point out the special difficulty of time-varying signal processing analysis caused by nonuniformity. As one outcome, it was found that the POCS method achieves a weighted pseudo-inverse of $S$ that outperforms the standard Moore-Penrose pseudo-inverse in terms of time-quantization noise reduction. But the broad framework analysis was also the opportunity to see what aspect of the discussed methods has a potential for generalizations. For example, the type of POCS method presented in this paper remains applicable with any integrating kernel function in place of the leakage function of LIF encoding. Although the POCS and Lazar's methods are defined on continuous-time signals, their iterations can be rigorously implemented in discrete-time in synchronization with the firing instants and without Nyquist-rate resampling, up to the use of a lookup table.

A point of investigation left for future research is the practical implementation of the POCS method by approximate time-varying FIR filters.

\appendix

\subsection{Proofs of Proposition \ref{prop:thn1} and Corollary \ref{prop:corol}}\label{app:innerprod}

\subsubsection{Preliminary}

\begin{lemma}
For any  function $ f(t)$,
\begin{eqnarray}\label{thn}
&(h_m* f)(t)=e^{-\alpha\Delta t_m}\,\ell(t{-}t_{m-1})-\ell(t{-}t_m)
\end{eqnarray}
where
$$\ell( t):=\displaystyle\int_0^ t e^{\alpha(t-s)} f( s)\dif s.$$
\end{lemma}

\begin{IEEEproof}
It follows from \eqref{hn} that
\begin{align*}
(h_m* f)(t)&=\int_{t_{m-1}}^{t_m}e^{\alpha( s-t_m)} f(t- s)\dif s\\
&=\int^{t-t_{m-1}}_{t-t_m}e^{\alpha(t- s-t_m)} f( s)\dif s\\
&=e^{-\alpha\Delta t_m}\int_0^{t-t_{m-1}}e^{\alpha(t-t_{m-1}-s)} f( s)\dif s\\
&\quad-\int_0^{t-t_m} e^{\alpha(t-t_m-s)} f(s)\dif s.
\end{align*}
\end{IEEEproof}

\subsubsection{Proof of Proposition \ref{prop:thn1}}

We have
$$\langle h_m, f_n\rangle=(h_m*\hat f)(t_n)\quad\mbox{where}\quad\hat f(t):= f(-t).$$
By applying \eqref{thn} and \eqref{Tmn}, we obtain
\begin{align}
\big\langle h_m, f_n\big\rangle&=
e^{-\alpha\Delta t_m}\ell(T^n_{m-1})-\ell(T^n_m)\label{thn3}\\
\mbox{where}\qquad \qquad
\ell( t)&:=\displaystyle\int_0^ t e^{\alpha(t-s)}\hat f(s)\dif s\nonumber\\
&=-\int_0^{-t}e^{\alpha(t+s)} f(s)\dif s=-g(-t)\qquad\nonumber
\end{align}
and $g(t)$ is given in \eqref{g-thn1}. Then, \eqref{thn3} leads to \eqref{thn1}.

\ppnoi
\subsubsection{Proof of Corollary \ref{prop:corol}}

From \eqref{thn}, we have
\begin{eqnarray*}
&\widetilde h_n(t)=(h_n*\varphi)(t)=e^{-\alpha\Delta t_n}\ell_{n-1}(t)-\ell_n(t)\\
&\hspace{-3.5mm}\mbox{where}\quad\ell_n(t):=\ell(t-t_n)\quad\mbox{and}\quad
\ell(t):=\displaystyle\int_0^ t e^{\alpha(t-s)}\varphi( s)\dif s,
\end{eqnarray*}
so that
$\big\langle h_m,\widetilde h_n\big\rangle=
e^{-\alpha\Delta t_n}\big\langle h_m,\ell_{n-1}\big\rangle-\big\langle h_m,\ell_n\big\rangle.$
It then follows from \eqref{thn1} that
\begin{align*}
&\big\langle h_m,\widetilde h_n\big\rangle
=e^{-\alpha\Delta t_n}\big(g(T^m_{n-1})-e^{-\alpha\Delta t_m}g(T^{m-1}_{n-1})\big)\\
&\hspace{2.5cm}-\big(g(T^m_n)-e^{-\alpha\Delta t_m}g(T^{m-1}_n)\big)\\
&\mbox{where}\qquad g(t):=\displaystyle\int_0^ t e^{\alpha(s-t)}\ell(s)\dif s\\
&=\int_0^ t e^{\alpha(2s-t)}\displaystyle\int_0^s e^{-\alpha \tau}\varphi( \tau)\dif \tau\dif s\\
&\textstyle=\left[\frac{e^{\alpha(2s-t)}}{2\alpha}\int\limits_0^ s e^{-\alpha\tau}\varphi(\tau)\,\dif\tau\right]_{ s=0}^t
\!\!\!-\int\limits_0^t\frac{e^{\alpha(2s-t)}}{2\alpha} e^{-\alpha s}\varphi( s)\,\dif  s\\
&=\midfrac{1}{2\alpha}\int_0^t e^{\alpha(t-\tau)}\,\varphi(\tau)\,\dif\tau  -\midfrac{1}{2\alpha}\int_0^t e^{-\alpha(t-s)}\,\varphi( s)\,\dif  s\\
&=\midfrac{1}{\alpha}\int_0^t\mathrm{sinh}(\alpha(t-s))\,\varphi( s)\,\dif  s.
\end{align*}

\bibliographystyle{ieeetr}

\bibliography{reference}{}

\end{document}